\documentclass[a4paper,11pt]{article}
\pdfoutput=1 

\usepackage{jcappub} 

\usepackage[T1]{fontenc} 

\usepackage{amsfonts}
\usepackage{amsmath}
\usepackage{amssymb}
\usepackage{graphicx,color}
\usepackage{float}
\usepackage{hyperref}
\usepackage{subfigure}
\usepackage{dcolumn}
\usepackage{soul}
\usepackage{ulem}
\usepackage{verbatim}
\usepackage{bm}  

\title{\texttt{WI2easy}: warm inflation dynamics made easy}

\author[a]{Gabriel S. Rodrigues}
\author[a,1]{and Rudnei O. Ramos \note{Corresponding author.}}

\affiliation[a]{Departamento de F\'{\i}sica Te\'orica, Universidade do Estado do Rio de Janeiro, 
20550-013 Rio de Janeiro, RJ, Brazil}

\emailAdd{gabriel.desenhista.gr.gr@gmail.com}
\emailAdd{rudnei@uerj.br}

\abstract{

We present \texttt{WI2easy}, a \texttt{Mathematica} package for high-precision analysis of warm inflation (WI) dynamics, 
enabling efficient computation of both background evolution and curvature perturbations. Designed with a user-friendly 
interface, the tool supports a broad spectrum of inflaton potentials--including large-field, small-field, 
and hybrid models--and accommodates arbitrary dissipation coefficients dependent on temperature, field amplitude, 
or both, encompassing canonical forms prevalent in WI studies. Users can define custom models through intuitive commands, 
generating full dynamical trajectories and perturbation spectra in a streamlined workflow. This facilitates rapid 
confrontation of theoretical predictions with observational constraints, empowering systematic exploration of 
WI parameter spaces. \texttt{WI2easy} bridges the gap between theoretical models and observational cosmology, 
offering a robust, adaptable framework for next-generation inflationary analyses.

}

\begin{document}
\maketitle

\clearpage
\section*{Program Summary}
\noindent
{\it Program title:} \texttt{WI2easy} \\[0.5em]
{\it Version:} 1.0  \\[0.5em]
{\it Program obtainable from:} \url{https://github.com/RudneiRamos/WI2easy} \\[0.5em]
{\it Distribution format:} .m \\[0.5em]
{\it Programming language:} \texttt{Mathematica} \\[0.5em]
{\it Computer:} Tested on a personal computer (CPU AMD Ryzen 9 3950X, 64 GiB RAM)\\[0.5em]
{\it Operating system:} Tested on Linux Mint (22.1) and Windows (10). \\[0.5em]
{\it Running time:} Depends on the complexity of the problem and code command used (see example notebook for typical runtimes).\\[0.5em]
{\it Nature of the problem:}  Analysis of warm inflation dynamics, enabling efficient computation of both background evolution and curvature perturbations.\\[0.5em]
{\it Solution method:}  Warm inflation perturbations solved using a deterministic {}Fokker-Planck formalism, bypassing the direct numerical solution of stochastic differential equations.\\[0.5em]
{\it Restrictions:} \texttt{Mathematica} version 13.3 or above. \\[0.5em]

\clearpage

\section{Introduction}

Warm inflation (WI)~\cite{Berera:1995wh,Berera:1995ie,Berera:1996nv,Berera:1996fm,Berera:1998px} is a cosmological framework in which the inflaton field continuously dissipates energy into a thermal radiation bath during the accelerated expansion phase, maintaining a non-negligible temperature throughout. Unlike standard cold inflation (CI), where the universe is supercooled and a reheating phase is required after inflation ends, WI naturally generates radiation through dissipative interactions (e.g., with other quantum fields), avoiding the need for a separate reheating epoch (for earlier reviews on the WI construction, see the refs.~\cite{Berera:2008ar,BasteroGil:2009ec}, while for a more recent
review see, e.g.~ref.~\cite{Kamali:2023lzq} and for a historical perspective on the development of the ideas in WI, see, e.g.~ref.\cite{Berera:2023liv}).
Key results in WI include modified density perturbations dominated by thermal (rather than quantum) fluctuations, leading to a distinct power spectrum with the possibility of a redder spectral tilt $n_s$ and suppressed tensor-to-scalar ratio $r$, aligning better with observational constraints for a larger variety of primordial inflation potential, including the simpler
monomial potentials~\cite{Bartrum:2013fia,Benetti:2016jhf,Bastero-Gil:2016qru}. Additionally, dissipative effects in WI can stabilize flat potentials against quantum corrections, mitigating the eta problem~\cite{Berera:1999ws,BasteroGil:2009ec,Bastero-Gil:2019gao} and predict unique signatures like non-Gaussianities~\cite{Bastero-Gil:2014raa,Mirbabayi:2022cbt} or running spectral indices~\cite{Das:2022ubr}, offering testable departures from CI paradigms. This interplay between thermodynamics and inflationary dynamics provides a bridge to particle physics models, such as those involving axions or strong coupling regimes, while addressing longstanding fine-tuning challenges.
It is fair to say that WI produces a
promising inflation picture that has been shown to better align with effective field theories with an ultraviolet (UV) completion in quantum
gravity~\cite{Das:2018rpg,Motaharfar:2018zyb,Goswami:2019ehb,Berera:2019zdd,Kamali:2019xnt,Berera:2020iyn,Das:2020xmh,Brandenberger:2020oav,Motaharfar:2021egj}.
These are all very appealing features that make WI a compelling scenario and which has been attracting increasingly interest in the recent literature.  

Since the beginning of the studies in WI, it was recognized that there are two important regimes for the dynamics during WI, the weak and strong dissipation ones, 
which can be characterized in terms of the ratio of the strength of the dissipation coefficient $\Upsilon$ over the Hubble parameter $H$,
$Q= \Upsilon/(3H)$, with the weak and strong regimes defined whether $Q$ is smaller or larger than one. Earlier detailed numerical analysis of the background dynamics
in these two regimes include for instance ref.~\cite{deOliveira:1997jt}, while studies of the perturbations in WI were done initially in ref.~\cite{Berera:1995wh} for the weak dissipative regime, and then in ref.~\cite{Berera:1999ws} for the strong dissipative regime. Simple approximate analytical expressions for the power spectrum for the perturbations in WI were derived, each one
valid in one of those two regimes. Since then, many papers in WI  made use of those simple approximated expressions, unfortunately not always
paying attention to the regime of validity  of them and sometimes extrapolating to regimes where they would not be valid. The first preliminary numerical study 
of the perturbations in WI was done in ref.~\cite{Taylor:2000ze}. Subsequently, a better estimate of the scalar of curvature 
power spectrum in WI, both from an analytical and numerical perspective was performed in ref.~\cite{Hall:2003zp}. 

The first approximate analytical expression
for the power spectrum in WI and valid in both weak and strong dissipative regimes was presented in ref.~\cite{Ramos:2013nsa}, which is still in use nowadays.
However, the result obtained in ref.~\cite{Ramos:2013nsa} did not take into account the coupling of the inflaton perturbations with the ones from the
radiation bath. {}For any typical dissipation coefficient that is a function of the temperature, as naturally expected in general in WI, and derived from many 
microscopic derivations~\cite{Gleiser:1993ea,Berera:1998gx,BasteroGil:2010pb,BasteroGil:2012cm,Bastero-Gil:2016qru,Bastero-Gil:2019gao,Berghaus:2019whh},
will naturally make the full perturbations in WI a coupled system. It was only in ref.~\cite{Graham:2009bf} that was fully appreciated that this coupling
of the inflaton and radiation perturbations would produce a profound effect on the power spectrum in WI. It was shown that dissipation coefficients
with a positive power in the temperature would lead to a growing mode for the perturbations as a function of $Q$, while a negative power in the temperature dependence of
the dissipation coefficient would produce a decreasing effect. These effects have been fully confirmed in refs.~\cite{Bastero-Gil:2011rva,Bastero-Gil:2014jsa} and which improved on the analysis done in ref.~\cite{Graham:2009bf}.
This effect of the coupling between the perturbations in WI is popularly
parameterized in terms of a function $G(Q)$ multiplying the analytical expression derived in the ref.~\cite{Ramos:2013nsa}. Unfortunately, so far we do not know any
analytical expression for this function $G(Q)$ and the existing results come from fittings of the full numerical power spectrum taking as a proxy the analytical
expression of ref.~\cite{Ramos:2013nsa}. Different fittings have been proposed already in refs.~\cite{Graham:2009bf,Bastero-Gil:2011rva}, with variations
for the $G(Q)$ function appearing subsequently, e.g. in refs.~\cite{Motaharfar:2018zyb,Kamali:2019xnt,Das:2020xmh,Mirbabayi:2022cbt}. These fittings for the $G(Q)$ function
have been of common use in many papers, unfortunately, once again, not always with an attention to the expected regime of validity of them or even using them for
forms of dissipation coefficients where they would not been designed for. This has been a serious matter for both producing results that can be reproducible and for implementing new
studies in WI. 
One important step tackling the above mentioned issues came with the release of the \texttt{WarmSPy} code~\cite{Montefalcone:2023pvh}.
The \texttt{WarmSPy}, a code written in Python language,  was the first development of a public code for WI specifically designed to
numerically  produce the function $G(Q)$ once a potential and dissipation coefficient are defined. 

In this paper, we present \texttt{WI2easy}, a \texttt{Mathematica} package designed to advance WI analyses by unifying the computation of perturbation equations and background dynamics. The code extends beyond conventional $G(Q)$ determinations by solving the full system of WI perturbation equations, while simultaneously evolving the inflationary background to self-consistently normalize the primordial power spectrum. This enables direct derivation of observational signatures--including the tensor-to-scalar ratio ($r$), spectral tilt ($n_s$), and scale-dependent curvature power spectrum 
$P_{\cal R}(k)$--critical for confronting theoretical predictions with observational datasets (e.g., Planck, BICEP/Keck). By automating the pipeline from model input (arbitrary potentials $V(\phi)$ and dissipation coefficients $\Upsilon(T\phi)$) to observables, \texttt{WI2easy} streamlines statistical validation of WI scenarios, empowering precision cosmology in the era of high-precision cosmic microwave background (CMB) and large-scale structure surveys.

\texttt{WI2easy} distinguishes itself from \texttt{WarmSPy} in its treatment of perturbation equations. While \texttt{WarmSPy} employs stochastic Langevin equations~\cite{Graham:2009bf,Bastero-Gil:2011rva,Kamali:2023lzq}, \texttt{WI2easy} adopts a deterministic {}Fokker-Planck formulation and which was recently proposed
in refs.~\cite{Ballesteros:2022hjk,Ballesteros:2023dno} -- a mathematically equivalent but computationally superior framework for deriving the scalar curvature power spectrum. This approach offers three key advantages:
(a) implementation simplicity -- the {}Fokker-Planck equations integrate naturally with \texttt{Mathematica}'s 
symbolic and numerical architecture, avoiding the algorithmic complexity of stochastic simulations;
(b) computational efficiency -- deterministic evolution of the coupled {}Fokker-Planck and background equations eliminates the need for ensemble averaging, reducing runtime by orders of magnitude;
(c) numerical precision -- direct access to \texttt{Mathematica}'s high-precision solvers ensures robust convergence, even for stiff systems or extreme parameter regimes.

By unifying the background and perturbation evolution within a deterministic framework, \texttt{WI2easy} streamlines precision analyses of WI, enabling direct confrontation with observational data without sacrificing computational tractability.
We can also take full advantage of \texttt{Mathematica} ecosystem: its symbolic engine automates derivation and simplification of dissipative dynamics, reducing manual errors, while its adaptive numerical solvers efficiently tackle stiff, nonlinear systems across parameter regimes (e.g., weak vs. strong dissipation). The platform's visualization tools further streamline analysis, enabling immediate plotting of field trajectories, power spectra and parameter dependencies to rapidly test observational viability against the observations data, facilitating the constraint of model
parameters. By unifying symbolic, numerical, and graphical workflows in a reproducible notebook format, \texttt{WI2easy} bridges analytical rigor with computational scalability, fostering collaborative, transparent exploration of WI
dynamics. This end-to-end integration positions the code as a critical tool for prototyping models in WI and that we hope will help to advance the study of WI at large.

This paper is organized as follows. In section~\ref{sec2}, we briefly
review the background equations in WI and give their implementation in the code. 
In section~\ref{sec3}, we present the full set of perturbation equations
in WI, the different choices of terms that can be added to them and
our choice of gauge. We also briefly review the {}Fokker-Planck implementation
for deriving the power spectrum in WI.
In section~\ref{sec4},
we describe in details all the functionalities of \texttt{WI2easy} and how to use it in practice.
In section~\ref{sec5}, we give several numerical results representative of some of the most
common WI models studied in the literature.
{}Finally, our concluding remarks are presented in section~\ref{conclusions}. One appendix is included
where we give more details about the equations entering in the code.

\section{The background equations in WI}
\label{sec2}

During WI, the inflaton is continuously decaying into radiation (with energy density $\rho_r$), which leads to
a thermal radiation bath with temperature defined by 
\begin{equation}
\rho_r = \frac{\pi^2 g_{*}}{30} T^4,
\label{rhorT}
\end{equation}
with $g_*$ denoting 
the effective number of relativistic degrees of freedom of the radiation bath. The WI regime is typically
characterized by the condition of having $T>H$, in which case, thermal effects due to the radiation bath
can lead to significant changes compared to the CI case.  The inflaton and radiation
energy density form a coupled system described by the dynamical system
\begin{eqnarray}
&&\ddot \phi + 3H(1+Q)\dot\phi + V_{,\phi}=0,
\label{phiEOM}
\\
&&\dot \rho_r + 4H\rho_r= 3HQ\dot \phi^2,
\label{radEOM}
\end{eqnarray}
where $V_{,\phi} = dV(\phi)/d\phi$, $H$ is the Hubble parameter ($M_{\rm Pl} \simeq 2.44\times 10^{18}$ GeV is the reduced Planck mass),
\begin{equation}
H^2 = \frac{1}{3 M_{\rm Pl}^2} \left( \frac{\dot \phi^2}{2} + V + \rho_r \right),
\label{hubble}
\end{equation}
and $Q$ is the dissipation ratio in WI,
\begin{equation}
Q= \frac{\Upsilon}{3H},
\label{eqQ}
\end{equation}
where $\Upsilon$ is the dissipation coefficient in WI. 
$Q$ essentially gives a measure of the strength
of the dissipative particle production effects in comparison to the
spacetime expansion. It also allows to characterize two regimes popularly studied
in WI: the $Q<1$ (weak dissipative regime) and $Q > 1$ (strong dissipative regime).
Each one of these regimes have differences with respect to both background and perturbation 
dynamics and have been studied extensively in the literature.
The dissipative processes in WI depends strongly on the dissipation coefficient $\Upsilon$.
Successful model realizations of WI based on microscopic physics (for a review,
see, e.g.~ref.~\cite{Kamali:2023lzq}), typically lead to forms of dissipation coefficients that are in general a function of both temperature
and the amplitude of the inflaton field, $\Upsilon \equiv \Upsilon(T,\phi)$. 

During the slow-roll inflationary regime, the eqs.~(\ref{phiEOM}), (\ref{radEOM}) and (\ref{hubble}) can be approximated and we can write them as
\begin{eqnarray}
 \dot\phi &\simeq &  - \frac{V_{,\phi}}{3H(1+Q)},
\label{phisr}
\\
\rho_r &\simeq& \frac{3}{4}Q\dot \phi^2\simeq \frac{3}{4}Q \left[ \frac{V_{,\phi}}{3H(1+Q)} \right]^2,
\label{radsr}
\end{eqnarray}
and
\begin{equation}
H \simeq \sqrt{\frac{V}{3 M_{\rm Pl}^2} }.
\label{hubblesr}
\end{equation}
The eqs.~(\ref{phisr}), (\ref{radsr}) and (\ref{hubblesr}) define the slow-roll trajectory
in WI. They also show that once initial values for $\phi$ and $Q$ are given (for fixed values
of the parameters in the inflationary potential), then $\dot \phi$ and $\rho_r$ 
are determined. This helps to set appropriate initial conditions and we use this in the code to generate 
the initial conditions once an initial value of $Q$ is given. Given an initial value $Q_i$, our 
algorithm solves the full background equations searching for the initial value of the inflaton,
$\phi_i$, that gives a specified number of e-folds of duration of the inflationary phase, $N_{\rm tot}$
(typically ranging between 50 and 60 {\it e}-folds). The initial conditions for the time derivative of the
inflaton and the radiation energy density then follows from the above relations,
$\dot \phi_i \equiv \dot \phi (\phi_i, Q_i, N_{\rm tot})$ and $\rho_{r,i} \equiv \rho_r (\phi_i, Q_i, N_{\rm tot})$,
with the end of inflation determined by when the accelerate regime for the scale factor ceases,
determined by the condition on the slow-roll coefficient, $\epsilon_H =1$, where
\begin{equation}
\epsilon_H = - \frac{\dot H}{H^2} \simeq \frac{\epsilon_V}{1+Q},
\label{epsH}
\end{equation}
with
\begin{equation}
\epsilon_V = \frac{ M_{\rm Pl}^2 }{2} \left( \frac{ V_{,\phi} }{V} \right)^2.
\label{epsV}
\end{equation}
At the same time, by writing the dissipation coefficient as a general function of temperature and 
the background inflaton field as
\begin{equation}
\Upsilon = C_{\Upsilon} f(T,\phi),
\label{Ups}
\end{equation}
we find the overall constant $C_\Upsilon$ in the dissipation coefficient as given by\footnote{We find the
procedure of first fixing $Q_i$ and then finding $C_\Upsilon$ much more convenient than the opposite.
This is because in WI the dissipation ratio $Q$ is a key quantity. Once appropriate ranges of
$Q$ for a given model are found, e.g. by requiring the results to be consistent with the observational data,
then automatically this will determine the corresponding values for $C_\Upsilon$. {}For any 
implementation of WI from first principles, the microscopic information about the model interactions 
leading to the dissipation coefficient derivation (e.g. coupling constants
and other parameters of the model), appears in $C_{\Upsilon}$ and can be constrained by its value.}
\begin{equation}
C_\Upsilon = \frac{3 H(\phi_i) Q_i}{f(T_i,\phi_i)},
\label{CUps}
\end{equation}
where $T_i$ is found from the radiation energy density of the thermal bath, eq.~(\ref{rhorT}).

\section{The perturbation equations in WI}
\label{sec3}

In WI, we have the perturbations from the inflaton, the radiation and the metric.
As shown in refs.~\cite{Graham:2009bf,Bastero-Gil:2011rva,Bastero-Gil:2014jsa}, these are described by coupled
stochastic equations and with no known simple analytical solution. However, by focusing on the
inflaton perturbation equation, a simplified solution can be found. This is possible, however, only after applying 
many approximations, e.g., dropping all slow-roll terms,
neglecting the metric perturbations and assuming that $Q$ is a constant. Under these simplify conditions, then both the homogeneous and
particular solutions for the equation can be determined. The resulting scalar of curvature
power spectrum under these approximations is found to be given by~\cite{Ramos:2013nsa}
\begin{eqnarray}
{P_{\cal R}}\bigr|_{analytic}&\simeq&
\frac{H_{*}^3T_{*}}{4\pi^2 \dot{\phi_{*}}^2}  \left[
  \frac{6 Q_{*} 2^{3Q_*}  \Gamma\left[ \frac{3(1+Q_*)}{2} \right]^2
\Gamma\left(\frac{3Q_*+1}{2}\right) }{ \Gamma\left( 1+ \frac{3Q_*}{2}\right) \Gamma\left(3Q_* + \frac{7}{2}
\right)} \right.  +
  \left.  \frac{H_{*}}{T_{*}}\left(1+2n_* \right)
  \right]
\nonumber \\
&\simeq& \left(\frac{H_*^2}{2\pi\dot\phi_*}\right)^2 \left(1+2n_*+\frac{2\sqrt3\pi
  Q_*}{\sqrt{3+4\pi Q_*}}\frac{T_*}{H_*}\right),
\label{PR1}
\end{eqnarray}
where
$\Gamma(x)$ is the Gamma-function,  the subindex $*$ means that
all quantities are to be evaluated at the Hubble crossing time, e.g.,
at the moment where the comoving mode satisfies $k=a H$, and
$n_*$ accounts for the possibility of a thermal distribution of the inflaton field due to
the presence of the radiation bath. When the inflaton perturbations are fully thermalized, $n_*$ assumes the
Bose-Einstein distribution, $n_* \equiv n_{\rm BE} = 1/[\exp(H/T)-1]$ (for intermediate cases,
see, e.g. ref.~\cite{Bastero-Gil:2017yzb}). 
Here we must note that the presence or not of this thermal contribution term is controversial in the literature.
In principle, one would expect that its presence or not should be strongly dependent 
on the details of the microphysics of the specific WI model under study, which would involve, for example, whether scattering rates of the inflaton
with other fields happen at a rate $\Gamma_{\rm scat}$ fast enough compared to the expansion rate, $\Gamma_{\rm scat} > H$,
such as to ensure thermalization of the inflaton\footnote{We note that \cite{Mirbabayi:2022cbt} has also presented arguments 
about the absence of this thermal contribution to the power spectrum, even when the inflaton thermalizes. 
However, the presence of the thermal bath
in WI might  give origin to other type of corrections to the scalar and tensor spectra 
that are of order ${\cal O} (H^3/T^3)$, as shown in \cite{Mirbabayi:2024eml}.}. 
We believe that the presence or not of this term can only be settled after an appropriate lattice simulation of the full dynamics,
or dedicated kinetic Botzmann equation study for the occupation number for the inflaton in WI is performed.
While this dedicated analysis is still lacking, most applications in WI
typically considered both possibilities (non thermalized inflaton perturbations, $n_*=0$, or fully
thermalized, $n_*=n_{\rm BE}$) and, hence, the results are presented when considering each one of them.
In our code, we give the option to include or not this term and users of the code can decide what would be more
appropriate for their analysis.

While we do not expect eq.~(\ref{PR1}) to match the full numerical result for the power spectrum in WI, 
we can overcome this shortcoming by using a common practice~\cite{Graham:2009bf,Bastero-Gil:2011rva,Bastero-Gil:2014jsa},
which is to solve the complete set of coupled perturbation equations and correct the analytical expressions eq.~(\ref{PR1}) by a
function $G(Q)$, derived from the numerical solution of the full coupled perturbation equations. The function
$G(Q)$ is then determined through a numerical fitting,
obtained by comparing the numerical power spectrum with the analytical proxy one, eq.~(\ref{PR1}), 
\begin{equation}
G(Q)=\frac{P_{\mathcal{R}}\bigr|_{numerical} }{P_{\mathcal{R}}\bigr|_{analytic}}.
\label{powersG}
\end{equation}
A systematic way of numerically deriving  $G(Q)$ for different inflaton potentials and dissipation coefficients
has been recently presented by the authors of ref.~\cite{Montefalcone:2023pvh} with the public code \texttt{WarmSPy}, a code
written in Python. Later, in Sec.~\ref{sec5}, we validate our code by comparing our results with the ones from the  \texttt{WarmSPy} and also with previous proposed fitting functions for $G(Q)$.

To obtain $G(Q)$, we make use of the complete set of perturbations equations. {}First, we present the relevant
metric perturbations. {}For that, we follow closely the notation and definitions of refs.~\cite{Hwang:1991aj,Hwang:2001fb}.
The  perturbed {}Friedmann-Lema\^{\i}tre-Robertson-Walker (FLRW) metric  is 
\begin{eqnarray}
ds^2 &=& -(1+2 \alpha) dt^2 - 2 a \partial_i \beta dx^i dt   \nonumber
\\ &+& a^2 [ \delta_{ij} (1 +2 \varphi) + 2 \partial_i \partial_j
  \gamma] dx^i dx^j, \label{metric}
\end{eqnarray}
with perturbations $\alpha$, $\beta$, $\gamma$ and $\varphi$. They 
are related to the complete set of metric equations
(working from now on in
space-momentum) through the definitions~\cite{Hwang:1991aj,Hwang:2001fb}
\begin{eqnarray}
&& \chi = a ( \beta + a \dot \gamma) \,, 
\label{chi} \\
&&\kappa= 3 (H \alpha - \dot \varphi) + \frac{k^2}{a^2} \chi \,, 
\label{kappa} \\
&&-\frac{k^2}{a^2} \varphi + H \kappa = - \frac{1}{2 M_{\rm Pl}^2}
\delta \rho\,,
\label{varphi}\\
&& \kappa -\frac{k^2}{a^2} \chi = - \frac{3}{2 M_{\rm Pl}^2} \Psi\,,
\label{kappachi}\\
&&\dot{\chi} + H \chi -\alpha -\varphi = 0\,,
\label{dotchi}\\
&&\dot{\kappa} + 2 H \kappa + \left(3 \dot{H} - \frac{k^2}{a^2}
\right) \alpha = \frac{1}{2 M_{\rm Pl}^2} \left( \delta \rho + 3\delta
p \right)\,,
\label{dotkappa}
\end{eqnarray}
where $k$ is the comoving wavenumber, 
$\delta \rho$, $\delta p$ and $\Psi$ are, respectively, the total density, pressure and momentum perturbations.
{}For our two-fluid equivalent system of inflaton field and radiation energy density, we have that 
\begin{eqnarray}
&&\delta \rho = \delta \rho_\phi + \delta\rho_r \,,
\label{deltarho}\\
&& \delta p = \delta p_\phi + \delta p_r\,,
\label{deltap}\\
&& \Psi = \Psi_\phi + \Psi_r\,,
\label{Psi}
\end{eqnarray}
with 
$\delta \rho_\phi = \dot{\phi} \delta\dot{\phi} - \dot{\phi}^2
\alpha + V_{,\phi} \delta \phi$, $\delta p_\phi = \dot{\phi}
\delta\dot{\phi} - \dot{\phi}^2 \alpha - V_{,\phi} \delta \phi$,
$\delta p_r = \omega_r \delta \rho_r$ and $\Psi_\phi = - \dot{\phi}
\delta \phi$, where $w_r$ is the radiation equation of state (which for a thermalized
radiation bath assumes the usual value $w_r=1/3$). Here ``dot'' always meaning  derivative with respect
to the cosmic time. In terms of these gauge-ready variables, the full evolution 
equations for the perturbation in the inflaton field, $\delta \phi$,
for the radiation perturbation, $\delta \rho_r$, and for the radiation momentum
perturbation, $\Psi_r$, are given by~\cite{Bastero-Gil:2011rva,Bastero-Gil:2014jsa}
\begin{eqnarray}
\delta \ddot{\phi} &+& 3H\left(1+Q\right)\delta \dot{\phi} +
\left(\frac{k^{2}}{a^{2}}+V_{,\phi\phi}+\frac{3 p
  HQ\dot{\phi}}{\phi}\right)\delta \phi  \nonumber \\ 
&+&
\frac{c H}{\dot{\phi}}\delta \rho_{r}- \dot{\phi}(\kappa+\dot{\alpha})
- [2\ddot{\phi}+3H(1+Q)\dot{\phi}]\alpha  =(6 QT)^{1/2} \xi^{(\phi)} ,  \nonumber \\
\label{deltaddotphi} 
\\ \delta \dot{\rho_{r}} &+& H \left(4 - \frac{3c Q\dot{\phi}^2}{4
  \rho_r}  \right)\delta \rho_{r} - \frac{k^2}{a^2}\Psi_{r} -
6HQ\dot{\phi} \delta \dot{\phi}  \nonumber \\
&-& \frac{3 p
  HQ\dot{\phi^{2}}}{\phi} \delta \phi - \frac{4  \rho_r}{3}  \kappa +
3 H\left( Q \dot{\phi}^2 + \frac{4\rho_r}{3}  \right) \alpha = -(6 QT)^{1/2} \dot \phi  \xi^{(\phi)},
\nonumber \\
\label{deltadotrhor} 
\\ \dot{\Psi}_{r}  &+& 3H \Psi_{r} + 3 H Q \dot{\phi} \delta \phi +
\frac{1}{3} \delta \rho_r   + 4 \rho_r \frac{\alpha}{3}=0 , 
\label{dotpsir}
\end{eqnarray}
where in the above equations, $c\equiv c(T,\phi)$ and $p\equiv p(T,\phi)$ are given by the logarithm derivative of the dissipation 
coefficient with respect to $T$ and to $\phi$, respectively,
\begin{eqnarray}
&&c=\frac{ \partial \ln \Upsilon(T,\phi)}{\partial \ln T},
\label{cvalue}
\\
&&p=\frac{ \partial \ln \Upsilon(T,\phi)}{\partial \ln \phi},
\label{pvalue}
\end{eqnarray}
and $\xi^{(\phi)}\equiv \xi^{(\phi)}({\bf k},t)$ is a stochastic Gaussian
source with two-point correlation function~\cite{Ramos:2013nsa}
\begin{eqnarray}
\!\!\!\!\!\!\!\!\!\!\langle \xi^{(\phi)}({\bf k},t)\xi^{(\phi)}({\bf
  k}',t')\rangle &=& \frac{1}{a^3} \delta(t-t') (2 \pi)^3
\delta({\bf k} + {\bf k}'). 
\label{noise1}
\end{eqnarray}
At this point, it is appropriate to comment on some the options offered by the code when concerning the above equations.
{}First, the presence of the stochastic noise appearing in the last term in the equation
for the radiation perturbation eq.~(\ref{deltadotrhor}), as first realized in ref.~\cite{Bastero-Gil:2014jsa}, 
is a consequence of  the conservation of the total stress energy tensor.
As discussed in ref.~\cite{Bastero-Gil:2014jsa}, its presence can be considered ambiguous since
the stress energy tensor can still be modified by the addition of appropriated stochastic energy flux terms, while
preserving its conservation.  Two possibilities have been explicitly shown in  ref.~\cite{Bastero-Gil:2014jsa},
e.g. leaving the stochastic term in the energy flux or in the momentum flux. Note also that most of
the previous works in WI just dropped this term in all the analysis of the perturbations 
(with some few exceptions, like in
refs.~\cite{Mirbabayi:2022cbt,Ballesteros:2022hjk,Ballesteros:2023dno}). The reason for that is that, as
shown in ref.~\cite{Bastero-Gil:2014jsa}, the effect of the noise term in eq.~(\ref{deltadotrhor}) is mostly
to affect the results in the weak dissipative regime of WI.
In our code implementation, we preserve the stochastic noise term in  eq.~(\ref{deltadotrhor}), but give the
option for the user to include or not it in the numerical simulations. Its effect can then be
clearly determined depending on the model under study.

A second point of relevance in our code implementation is the inclusion or not of the
term $n_*$ (which we set as in the fully thermalized case, $n_*=n_{\rm BE}$) in the power spectrum expression
eq.~(\ref{PR1}). As already mentioned, the presence or not of this terms is strongly dependent 
on the details of the microphysics of the specific WI model under study.
In the absence of the thermal distribution term, we can simply
solve the inflaton perturbation equation assuming standard Bunch-Davis initial conditions.
This ensures that the homogeneous part of the solution for $\delta \phi$ will reproduce the
CI result $\langle|\delta \phi|\rangle = H/(2\pi)$. When considering the presence of the thermal contribution
$n_{\rm BE}$, we make use of the proposal of ref.~\cite{Ramos:2013nsa}, of changing the perturbation 
equation for $\delta \phi$, eq.~(\ref{deltaddotphi}) and adding to the right-hand-side of it
another Gaussian stochastic noise term\footnote{Note that in this case, we can just set the initial
conditions for the inflaton perturbations to zero, as the extra noise term already ensures that
the correct CI result for the spectrum spectrum appears from the particular solution
as a consequence of the extra noise term.}, 
\begin{equation}
\left[H^2(9+12\pi Q)^{1/2}(1+2n_*)/\pi\right]^{1/2} \xi^{(q)},
\label{quantumnoise}
\end{equation}
where the noise term $\xi^{(q)}$ satisfies a two-point correlation function like (\ref{noise1})
and the amplitude of the noise in (\ref{quantumnoise}) is such that to correctly reproduce the
form of the power spectrum in eq.~(\ref{PR1}). 
Here, we avoid the prescription considered in ref.~\cite{Ballesteros:2023dno} of multiplying
the part of the spectrum coming from the homogeneous solution for $\delta\phi$ by a factor $(1+2 n_*)$ at the end.
The prescription adopted here concerning this thermal contribution to the power spectrum
is also the same one considered in the \texttt{WarmSPy} code (see ref.~\cite{Montefalcone:2023pvh}
for details of their implementation).
In our code, we also make available the option of including or not the thermal term $n_*$
in the numerical simulations.

Besides of the perturbation equations (\ref{deltaddotphi}), (\ref{deltadotrhor}) and (\ref{dotpsir}),
we must complement them with the equations for the metric ones once some gauge choice is made. 
Some of the choices of
gauges previously considered in the literature include the zero-shear
gauge~\cite{Bastero-Gil:2014jsa}, where $\chi=0$, the Newtonian gauge~\cite{Montefalcone:2023pvh,Ballesteros:2023dno},
or directly in terms of gauge invariant quantities~\cite{Bastero-Gil:2011rva}. In the present
work, we found that the use of the uniform-curvature gauge, where $\varphi=0$, to be a more suitable
choice for the gauge. This choice has advantages both from a numerical viewpoint and can also facilitate future planned extensions of the present code.
In the uniform-curvature gauge we can then find the relations
\begin{eqnarray}
 H\kappa&=&-\frac{1}{2 M_p^2} \left( \dot\phi \delta \dot\phi - \dot\phi^2 \alpha + V_{,\phi}\delta \phi + \delta \rho_r\right),
\label{metric1}
\\
H\alpha &=&\frac{1}{2 M_p^2} \left( \dot\phi \delta\phi -\Psi_r \right),
\label{metric2}
\\
H\dot\alpha &=& -\dot H \alpha +\frac{1}{2 M_p^2}\left\{\left[-3H(1+Q)\dot \phi - V_{,\phi}\right]\delta \phi + \dot\phi \delta \dot\phi - \dot\Psi_r\right\},
\label{metric3}
\end{eqnarray}
Thus, the set of equations (\ref{deltaddotphi})-(\ref{dotpsir}), (\ref{metric1})-(\ref{metric3}), form a closed system of
equations whose only perturbation quantities are $\delta \phi,\; \delta \dot\phi,\; \delta \rho_r$ and $\Psi_r$, which
together with the background equations (\ref{phiEOM}) and (\ref{radEOM}), can be solved numerically.
{}From the solution of the above equations, one finally obtain the power spectrum, which is defined as~\cite{Bastero-Gil:2011rva,Bastero-Gil:2014jsa}
\begin{equation}
P_{\cal R}(k)= \frac{k^3}{2 \pi^2} \langle |{\cal R}|^2
\rangle\,, \label{PR}
\end{equation}
where $\langle \ldots \rangle$ means here the ensemble average over the
different realization of the noise terms in the perturbation equations
and $\mathcal{R}$ is defined as (in the uniform-curvature gauge)
\begin{eqnarray}
&&{\cal R}= \frac{H}{\rho_T+p_T}
\left(\dot \phi \delta \phi - \Psi_r\right),
\label{calR}
\end{eqnarray}
where $\rho_T +p_T = \dot\phi^2 + 4\rho_r/3$,  using $p_{r} =\rho_{r}/3$.
As usual, the power spectrum $P_{\cal R}$ is evaluated at the time where the modes
freeze, $k \ll aH$.

In our implementation of the solution of these equations, instead of solving the stochastic equations as they stand
and done like e.g. in refs.~\cite{Bastero-Gil:2011rva,Bastero-Gil:2014jsa,Montefalcone:2023pvh}, we follow the
procedure put forward in the refs.~\cite{Ballesteros:2022hjk,Ballesteros:2023dno}. In the procedure of refs.~\cite{Ballesteros:2022hjk,Ballesteros:2023dno},
instead of solving the stochastic system of equations, one can write down the power spectrum in terms of the probability distribution
obtained from the Fokker-Planck equation derived from a matrix Langevin equation and defined from the system of stochastic differential
equations for the perturbations.
In essence, we define a vector with the perturbation quantities, ${\bm \Phi}$, which in the present case, is a four component (column) vector\footnote{Note that we solve 
the system of equations in terms of the number of {\it e}-folds instead of time variable, $dN_e = H dt$.}
\begin{equation}
{\bm \Phi} = \left(\delta \phi, \frac{d \delta \phi}{dN_e}, \delta \rho_r, \Psi_r\right)^T.
\label{Phi}
\end{equation}
Our system of equations (\ref{deltaddotphi}), (\ref{deltadotrhor}) and (\ref{dotpsir}) is then written down
as a coupled first-order Langevin equations with matrix form given by
\begin{equation}
\frac{d {\bm \Phi}}{dN_e} + {\bf A} {\bm \Phi} = {\bf B} \xi_{N_e},
\label{langevin}
\end{equation}
where ${\bf A}$ is here a $4\times 4$ matrix, ${\bf B}$ is a column matrix and $\xi_{N_e}$ is a Gaussian stochastic noise
with two-point correlation function given by
\begin{equation}
\langle \xi_{N_e}({\bf k}) \xi_{N_e'}({\bf k}')\rangle = \delta(N_e-N_e') (2 \pi)^3
\delta({\bf k} + {\bf k}').
\label{xiN}
\end{equation}
In terms of ${\bm \Phi}$, the power spectrum (\ref{PR}) can be written in matrix form as
\begin{equation}
P_{\cal R}(k)= \frac{k^3}{2 \pi^2} {\bf C}^T .\langle {\bm \Phi} {\bm \Phi}^\dagger 
\rangle . {\bf C}\,, \label{PRPhi}
\end{equation}
where ${\bf C}$ is another column matrix of four elements and which can be written from
eqs.~(\ref{PR}) and (\ref{calR}), and $ \langle {\bm \Phi} {\bm \Phi}^\dagger
\rangle$ is defined as
\begin{equation}
\langle {\bm \Phi} {\bm \Phi}^\dagger\rangle
=\int \prod_i d { \Phi}_i \int \prod_j d { \Phi}_j^* P({\bm \Phi},{\bm \Phi}^*,N_e) {\bm \Phi} {\bm \Phi}^\dagger,
\label{PhiPhi}
\end{equation}
with $P({\bm \Phi},{\bm \Phi}^*,N_e)$ is the probability density, which is the solution of the 
{}Fokker-Planck equation,
\begin{equation}
\frac{\partial P}{\partial N_e} = \sum_{i,j} \left[ A_{ij}\frac{\partial}{\partial \Phi_i} \left(\Phi_j P\right)
+A_{ij}\frac{\partial}{\partial \Phi_i^*} \left(\Phi_j^* P\right) + \left({\bf B} {\bf B}^T\right)_{ij}
\frac{\partial^2 P}{\partial \Phi_i \partial \Phi_j^*} \right].
\label{FP}
\end{equation}

In the {}Fokker-Planck approach of refs.~\cite{Ballesteros:2022hjk,Ballesteros:2023dno}, instead of solving
the Langevin equation~(\ref{langevin}) to obtain the power spectrum, we solve instead the  differential
equation for the  two-point statistical momenta, ${\bf J} \equiv  \langle {\bm \Phi} {\bm \Phi}^\dagger\rangle$,
which from eqs.~(\ref{PhiPhi}) and (\ref{FP}), it satisfies
\begin{equation}
\frac{d {\bf J}}{dN_e} = - {\bf A}.{\bf J} - {\bf J}.{\bf A}^T + {\bf B}.{\bf B}^T.
\label{Jequation}
\end{equation}

In the \texttt{WI2easy} code the solution of the differential eq.~(\ref{Jequation}) is implemented
in \texttt{Wolfram Mathematica} and its solution used to obtain the power spectrum eq.~(\ref{PRPhi}).
These equations are solved simultaneously with the background equations for WI.
Details of the form of the equations entering in the code, along also with the definitions
of the matrices ${\bf A},\; {\bf B}$ and ${\bf C}$ are given in the appendix~\ref{appA}.

In the next section we explain the code usage.

\section{The \texttt{WI2easy} code overview and usage}
\label{sec4}

\texttt{WI2easy} is a package of routines written in \texttt{Wolfram Mathematica} 
language that numerically computes the background and perturbation differential
equations in WI. It is released as a .m file and can be downloaded from
\url{https://github.com/RudneiRamos/WI2easy}.

The workflow of \texttt{WI2easy} is the following:

\begin{center}

\fbox{
    \begin{minipage}{\textwidth}
Define inflaton potential, dissipation coefficient and any parameters for them. Specify the
type of model (large field, small field or hybrid). Specify whether to consider thermalized
inflaton perturbations or not. Specify whether the add radiation noise term in the radiation perturbation
equation or not
    \end{minipage}
}

\bigskip

{\huge $\Downarrow$ }

\bigskip

\fbox{\mbox{Run code command to generate the initial conditions for the desired duration of inflation}} 

\bigskip

{\huge $\Downarrow$ }

\bigskip

\fbox{\mbox{Run code command to generate the function $G(Q)$ }} 

\bigskip

{\huge $\Downarrow$ }

\bigskip

\fbox{
    \begin{minipage}{\textwidth}
Run code command to produce the background evolution with proper CMB normalization  of
 the potential and to generate the power spectrum, $r$, $n_s$, etc (for a specified initial value of $Q$)
    \end{minipage}
}

\bigskip

{\huge $\Downarrow$ }

\bigskip

\fbox{
    \begin{minipage}{\textwidth}
Optional: run code command to generate several plots for the background
quantities and power spectrum
    \end{minipage}
}

\bigskip

{\huge $\Downarrow$ }

\bigskip

\fbox{\mbox{Run code command to produce main background quantities, $r$, $n_s$, etc as a function of $Q_*$}} 

\end{center}

\subsection{Installation}

\texttt{WI2easy} can be installed like any other \texttt{Wolfram Mathematica} package.
Just  place the package file WI2easy.m  in the applications folder \texttt{Mathematica/Applications}
or any other directory in the \texttt{Mathematica} path.
Alternatively, the easiest way (recommended) is just to put the WI2easy.m package file in any working 
directory, along also with the companion \texttt{Mathematica} notebook \texttt{main.nb} and
simple run the commands in that sample notebook. 

\subsection{Running \texttt{WI2easy}}

In the \texttt{main.nb} notebook, after loading \texttt{WI2easy} with the command
\texttt{<< WI2easy.m}, the next command  is \texttt{ ParametersWI[]}.

\subsubsection{Command \texttt{ ParametersWI[]}}

By running the  command \texttt{ ParametersWI[]}, a window will open and where all the relevant quantities can be input:

\begin{enumerate}
    \item Inflaton potential: write your potential (for example: \texttt{V0 $x^4/4$}, for
a quartic potential). It is mandatory to include the $V0$ normalization (or for the potential
to have a free constant named $V0$ that can be used later when normalizing it according to the
CMB amplitude of the scalar power spectrum at the pivot scale. Here $x$ is the inflaton field
normalized by the reduced Planck mass $M_{\rm Pl}$. 

    \item $V0$ normalization: an initial value for the normalization (in dimensionless units) must be
given (for example: \texttt{$10^{-14}$}, like for the quartic potential example). Do not worry about the
precise value of $V0$ here. It can be a value close to what would be obtained for CI.
The correct value of $V0$ will be computed later automatically by the code.

    \item Potential shape: specify which type of model you are working with, e.g. large field, small
field or hybrid model (in the example of the quartic potential, it is a large field potential)\footnote{
The classification of large or small field potentials is the usual from the literature.
Any potential where the inflaton starts at a large value and runs down to a minimum at $\phi=0$
is a large field model (e.g. monomial chaotic potentials, Starobinsky potential, $\alpha$-attractor potentials
in general, all fall in this category). Otherwise, if the inflaton field starts in a plateau around the origin
($\phi=0$) and runs towards large field values, it is a small field potential (e.g. hilltop potentials,
axion type of potentials, etc)}.
If the choice is a hybrid potential, then specify the critical value for the inflaton field, which determines
where inflation will end (do not add the waterfall field. This is still considered a one-field model).

 \item Potential name: give some simple name for your potential (e.g. \texttt{quartic}, in the case of the quartic
potential example). This will be used for generating the name of the data files.

    \item Initial inflaton field (in reduced Planck 
mass units): to be used in the search algorithm of the initial conditions, generally a value close to the one found for CI for the model under consideration (e.g. \texttt{25}, for the quartic inflaton potential example).

\item Functional dependence of the dissipation coefficient: write the dependence of the dissipation coefficient
in $T$ (temperature) and $x$ (the normalized inflaton field). Typical examples are: $T$, $T^3$, $T^3/x^2$, etc.

\item Dissipation name: give a simple name for the dissipation coefficient being used
(e.g. \texttt{linearT}). This will be used for generating the name of the data files.

\item Thermalized fluctuations ? Whether to include the thermal distribution in the spectrum and
perturbation equations or not (see explanation in the previous section). Answer Yes or No.

\item Radiation noise ? Whether to include the radiation noise term in the radiation
perturbation equation or not (see explanation in the previous section). Answer Yes or No.

\item Relativistic degrees of fredom (DoF): give the value for $g_*$ (  
examples: \texttt{106.75}, for the standard model, \texttt{228.75}, for the minimal supersymmetric model, etc).

\item Number of {\it e}-folds: Give the number of {\it e}-folds that you want inflation to last (e.g.  \texttt{60}).

\item Defining extra parameters: If the model (potential, dissipation, etc) has any dependent constants that you need to define, you can defined them here.

\end{enumerate}

Check the values and expressions given by running the command \texttt{pars}.
Note that all expressions and values given in the initial window opened by \texttt{ ParametersWI[]}
can be overwritten in the \texttt{ main.nb} notebook (see example in the provided notebook).

After defining the potential and dissipation coefficient and the basic running information about
the model. The user is set to run the next command available.

\subsubsection{Command \texttt{FindICs[Qi,Qf]}}
\label{secICs}

By running the  command \texttt{FindICs[Qi,Qf]}, the background equations are solved and the code
will run the search algorithm to find the appropriate initial conditions (ICs) for the model and that lead to the specified 
total duration in {\it e}-folds. The ICs will be generated for values of the dissipation ratio
in the interval $[Q_i,Q_f]$. The results will be saved in the data file starting with \texttt{ICs$\_$data}.
Here, another data file is also produced and saved with the name \texttt{cp$\_$data}, which contains the values of $c$ and $p$
defined by eqs.~(\ref{cvalue}) and (\ref{pvalue}), respectively. The results for $c$ and $p$ are 
used in subsequent parts of the code.

The command \texttt{FindICs} can also be used for a quick run of the code for the model specified and
for only one value of $Q$. {}For that, just set the values for $Q_i$ and $Q_f$ to be the same,
\texttt{FindICs[Q,Q]}. The available background functions after running the command are
$\phi$, $\dot \phi$, $\rho_r$, $Q$, $T/H$, $T$, $H$ and $\epsilon_H$ (all in units of $M_{\rm Pl}$)
as a function of the number of {\it e}-folds. They can be accessed through the functions
\texttt{phi[ne]}, \texttt{dphi[ne]}, \texttt{rad[ne]}, \texttt{Qdiss[ne]}, \texttt{TH[ne]}, \texttt{Temp[ne]}, \texttt{Hubble[ne]} and \texttt{epsilon0[ne]}, respectively. This can be useful for a quick check of the general
behavior of the background quantities for the model being studied. Note that at this stage all results are
produced with the inflaton potential normalization $V_0$ set at the initial configuration stage.

\subsubsection{Command \texttt{FindGQ[]}}

After generating the ICs in the previous stage, it can now be issued the  command \texttt{FindGQ[]}.
Here, the code will make use of the ICs in the interval $[Q_i,Q_f]$ set in the previous command, run the
perturbation equations using eq.~(\ref{Jequation}) and compute the power spectrum from eq.~(\ref{PRPhi}).
{}For each value of $Q \in  [Q_i,Q_f]$, it is used a value of scale $k$ in the perturbation
equations that corresponds to seven {\it e}-folds
before Hubble radius crossing (whose values are also computed when running the previous command \texttt{FindICs[Qi,Qf]}), which correspond to a value around $k \simeq 10^3 a(t_i) H(t_i)$).
The perturbations are then evolved an additional seven {\it e}-efolds, which ensures the spectrum 
has already been frozen and $k \ll  a H$. The obtained numerical power spectrum is then compared to the analytical proxy result
eq.~(\ref{PR1}) and the function $G(Q)$ computed from the definition eq.~(\ref{powersG}). 
The results are saved in the data file starting with \texttt{G(Q)$\_$data}.
Note that the power spectrum that is evaluated at this stage is still the non-normalized one.

\subsubsection{Command \texttt{ObservationsWI[Q0value,x0value]}}

Having generated the appropriate ICs and the results for $G(Q)$, by now issuing the command \texttt{ObservationsWI[Q0value,x0value]}, the value of $V_0$ for the normalization of the
potential will now be determined from the amplitude of the scalar spectrum at the pivot scale.
Here, it is assumed the amplitude for the scalar spectrum, $\ln\left(10^{10} A_s \right) = 3.047$ (from the Planck Collaboration~\cite{Aghanim:2018eyx}
TT,TE,EE-lowE+lensing+BAO 68$\%$ limits), at the pivot scale $k_{\rm pivot}=0.05 \, {\rm Mpc}^{-1}$. 
Both the amplitude of the scalar spectrum and/or the pivot scale can be set to other values in the package code
if needed. 

\texttt{ObservationsWI[Q0value,x0value]} will execute the following steps: (1) for the value
of \texttt{Q0value} for $Q$ and starting from the initial value \texttt{x0value} for the inflaton, the code will 
search for the appropriate conditions leading to the total number of {\it e}-efolds set at the beginning of
the code, but this time it will also make sure that the power spectrum at the initial number of {\it e}-folds will 
have the correct amplitude. This way, the correct value  $V_0$ for the normalization of inflaton potential
is determined. Here, the code makes uses of the proxy power spectrum eq.~(\ref{PR1}) but corrected by
the function $G(Q)$. Instead of using any numerical fitting for the data for $G(Q)$ generated 
from the previous command \texttt{FindGQ[]}, the code makes use of the raw data from the \texttt{G(Q)$\_$data} file
and does a spline interpolation of the data to create a function $G(Q)$ that can be used internally in the calculations.
At this stage, the code still makes use of the slow-roll conditions and the approach described at the end
of sec.~\ref{sec2}. The code performs then the second step. (2) To avoid any issues of numerical precision due to the use of the approximated equations (and e.g. the field trajectory not to be in the inflationary attractor one),
the code will run the background equations backwards for about seven {\it e}-efolds and run it again forward from that
point until the end of inflation. The point $N_*$ where the pivot scale crosses the Hubble radius (where the amplitude
of the power spectrum at the pivot scale is given by the value $A_s$), the point
where inflation ends $N_{\rm end}$ (i.e., where $\epsilon_H =1$) and also the point $N_{\rm reh}$ where the universe transits to the radiation
dominated regime after inflation (i.e., where $\epsilon_H \simeq 2$) are determined. 
The code also verifies whether the value of $N_*$ satisfies the
correct point where the pivot scale have crossed the Hubble radius, given by the condition~\cite{Das:2020xmh}\footnote{In practice, the code always check whether eq.~(\ref{N*})
is satisfied or not, adjusting $N_{tot}$ and the initial conditions until the condition given by eq.~(\ref{N*}) is fulfilled.
Although most of the evolution for the background
quantities is presented starting from the point $N_*$ (which follows the usual convention of always meaning  the number of e-folds before the
end of inflation for which the pivot scale crosses the Hubble radius), it is also useful to present results before this
point (e.g. for $N_{tot}>N_*$), like, for example, for the evolution of the power spectrum (see, e.g. fig.~\ref{fig7}(a) below). 
This way we can cover scales both below and above the pivot scale.}
\begin{equation}
\frac{k_p}{a_0 H_0} = e^{-N_*} \left[ \frac{43}{11 g_s(T_{\rm end})}
  \right]^{1/3} \frac{T_0}{T_{\rm end}} \frac{H_*}{H_0}  \frac{a_{\rm
    end}}{a_{\rm reh}},
\label{N*}
\end{equation}
where $g_s(T_{\rm end})$ in the entropy number of degrees of freedom at the end of inflation, which in the code is considered to the same as the value set for $g_*$\footnote{The difference between $g_s$ and $g_*$ is not expected to be relevant since
we typically deal with WI models where the energy scale at the end of inflation is around
and above $10^9$ GeV or so. The final results are also weakly dependent on the precise value of $g_*$.}. The  default pivot scale used is $k_p = 0.05/{\rm Mpc}$, $a_0=1$ is the scale factor today, $H_0=67.66\, {\rm km}\, s^{-1}
{\rm Mpc}^{-1}$ (from the Planck Collaboration~\cite{Aghanim:2018eyx},
  TT,TE,EE-lowE+lensing+BAO 68$\%$ limits) and $T_0 = 2.725\, {\rm K} \simeq 2.349
\times 10^{-13}\, {\rm GeV}$ is the present day value of the CMB temperature. In the above equation, 
$a_{\rm end}/a_{\rm reh}$ gives the duration (in $e$-folds) lasting from the end of inflation until the beginning
of the radiation dominated phase (determined by solving the background equations and tracking the point where 
the equation of state after WI becomes approximately 1/3). In WI, this transition period from the end of inflation to the start
of the radiation dominated regime typically lasts around $1-6$ {\it e}-folds.
If the condition (\ref{N*}) is not met, the code will repeat the previous steps until it converges
to the value obtained from eq.~(\ref{N*}) within $3\%$ precision, which for all cases we have studied suffices
when obtaining the observable quantities. At the final stage, the command code will evaluate 
the tensor-to-scalar ratio $r$,
\begin{equation}
r = \frac{P_{\cal T}}{P_{\cal R}} ,
\label{r}
\end{equation}
where $P_{\cal T} = 2 H^2/(\pi^2 M_{\rm Pl}^2)$, 
the 
spectral tilt $n_s$, 
\begin{equation}
n_s =   1+ \frac{d \ln P_{{\cal R}}(k/k_*) }{d \ln(k/k_*)}\Bigr|_{k\to k_*} ,
\label{eq:n}
\end{equation}
and the running of the spectral tilt\footnote{We do not evaluate the running of the running (see, e.g.
ref.~\cite{Das:2022ubr}) since the present CMB data does not put clear constraints on it (even the
most update data results for the running are also rather poor~\cite{ACT:2025tim}). Nevertheless, the
running of the running can be easily implemented and computed from the results obtained with the code.}
\begin{equation}
\alpha_s=   \frac{dn_s(k/k_*) }{d \ln(k/k_*) }\Bigr|_{k\to k_*}.
\label{dns}
\end{equation}

In addition to the background quantities accessible through the functions already described for the usage of the command \texttt{FindICs[Qi,Qf]} explained earlier, here we will also have access for the additional results for the spectrum $P_{\cal R}$, $r$, $n_s$ and $\alpha_s$.
These results are accessible through the respective commands: \texttt{spectrum[ne]}, \texttt{rtensor[ne]}, \texttt{ns[ne]}
and \texttt{dns[ne]}, respectively, all given in terms of the number of {\it e}-folds and that can be plotted
from the initial time ($N_e=0$) seven {\it e}-folds before $N_*$ (in the code it is called  \texttt{Nstar})
until the beginning of the radiation dominated regime $N_{\rm reh}$ (in the code it is called \texttt{Nre}, while
the end of inflation is given by the value \texttt{Nend}).

To make easier the visualization of the different results, after running this part of the code, 
with the command  \texttt{MakePlotsEvolution} several plots can be made.
\texttt{MakePlotsEvolution} will give plots showing the evolution of background quantities, starting at $N_*$, and also for the spectrum,
starting at $N_e=0$, such that we can cover a region of scales both below and above the pivot scale and
which are within the window of accessible scales with the CMB data.  

\subsubsection{Command \texttt{QrangeWI[Qinit]}}

The code command \texttt{QrangeWI[Qinit]} will repeat the execution of the previous command, varying $Q$
from an initial value $Q_i$ up to the largest value of $Q$ available from the data file \texttt{ICs$\_$data}
generated by the \texttt{FindICs} command. The command will stop in case $n_s$, for some value of $Q$, 
gets outside of the range $n_s \in [0.93,0.99]$. Results for the
background and perturbation quantities at the Hubble radius crossing $N_*$ for each value of $Q$ considered are
then saved in the data file \texttt{observables$\_$data}. Examples of several plots that can be made with
these results are given in the last section of the companion \texttt{Mathematica} notebook \texttt{main.nb}.

\section{Models analyzed} 
\label{sec5}

To illustrate the results that can be obtained from the code, we will focus on two cases
of large field models, namely:

\begin{itemize}

\item the quartic monomial potential,
\begin{equation}
V(\phi) = \frac{V_0}{4} \left(\frac{\phi}{M_{\rm Pl}} \right)^4 \equiv \frac{\lambda}{4} \phi^4,
\label{phi4}
\end{equation}
where $\lambda= V_0/M_{\rm Pl}^4$.

\item the fibre type I inflation potential~\cite{Cicoli:2008gp},
\begin{equation}
V(\phi) = V_0 \left[F-4 e^{-\frac{\phi}{\sqrt{3} M_{\rm Pl}}}+ e^{-4\frac{\phi}{\sqrt{3} M_{\rm Pl}}}+R \left(
e^{2\frac{\phi}{\sqrt{3} M_{\rm Pl}}}-1 \right) \right],
\label{fibre}
\end{equation}
where $F$ and $R$ are dimensionless constants. Here, we assume $R$ to have the value $R=3.05\times 10^{-7}$, which is motivated from
recent studies with this type of potential~\cite{Cicoli:2020bao,Bhattacharya:2020gnk,Bera:2024ihl}, while the value of $F$ is adjusted
such that at the minimum of the potential it satisfies $V(\phi_{\rm min})=0$.
This is a potential motivated from string theory constructions and has analogies with $\alpha$-attractor type of
potentials. A detailed analysis of fibre inflation in the context of WI is given in ref.~\cite{Chakraborty:2025yms}.

\end{itemize}

We also consider two cases of small field models, namely:

\begin{itemize}

\item a cosine axion-type of potential,
\begin{equation}
V(\phi) =V_0
\left[1+ \cos\left(\frac{\phi}{f_a}\right)\right],
\label{axion}
\end{equation}
where $f_a$ is the axion decay constant, which in our example we consider the value $f_a=5M_{\rm Pl}$,
like considered in ref.~\cite{Montefalcone:2023pvh} and which allows us to better compare our results
with the ones obtained in that reference.

\item a quadratic hilltop potential,
\begin{equation}
V_{\rm hilltop}(\phi) =V_0
\left[1-\gamma\left(\frac{\phi}{M_p}\right)^{2}\right],
\label{hilltop}
\end{equation}
where $\gamma$ is a constant, which here we take to have the value $\gamma=0.01$ (note that for $\phi/f_a \ll 1$,
the axion potential eq.~(\ref{axion}) reduces to a form like (\ref{hilltop}) with
the identification $\gamma = M_{\rm Pl}^2/(4 f_a^2) $).

\end{itemize}

{}Finally, we also give an example of a hybrid inflaton potential. 

\begin{itemize}

\item The traditional form of the
effective potential of hybrid inflation is a two-field model~\cite{Linde:1993cn}, 
composed of the inflaton field $\phi$ and a
waterfall field $\sigma$. The potential is
\begin{eqnarray}
  V(\phi,\sigma)=\frac{1}{4\lambda}(M^2-\lambda\sigma^2)^2+\frac12m^2\phi^2
  +\frac12g^2\phi^2\sigma^2.\nonumber\\
\label{Vhybrid}
\end{eqnarray}
{}From the above potential, during inflation the mass square of the $\sigma$ field is 
$m_\sigma^2(\phi)=-M^2+g^2\phi^2$. Inflation takes place when  $\phi>M/g$
and $\sigma=0$ is the only one minimum (true vacuum) in the $\sigma$-field direction.
Hence, while $\phi > \phi_c$, where the
critical value for the inflaton is
$\phi_c=M/g$, the potential is simply
\begin{eqnarray}
V_{\rm eff}(\phi)=\frac{M^4}{4\lambda}+\frac12m^2\phi^2,
\label{pothyb}
\end{eqnarray}
and the constant term $\frac{M^4}{4\lambda}$ $(\gg
\frac12 m^2\phi^2 )$ drives the expansion.  When $\phi$ gets below $\phi_c$,  
the waterfall field $\sigma$ rolls down to one of its
new minima $\sigma(\phi)=\pm M_\sigma(\phi)/\sqrt{\lambda}$ and inflation quickly 
terminates. In the code we write the potential~(\ref{pothyb}) as
\begin{eqnarray}
V_{\rm eff}(\phi)=\frac{M^4}{4\lambda}+\frac{V_0}{2}\frac{\phi^2}{M_{\rm Pl}^2},
\label{pothybV0}
\end{eqnarray}
and let the code automatically compute $V_0$ such as to give the proper CMB
amplitude at the pivot scale. {}For the numerical example for this potential shown in the next
section, we made the choice of parameters: $M=10^{-5} M_{\rm Pl}$, $\lambda=0.1$
and $g=0.05$. 

\end{itemize}

As dissipation coefficients, we consider three of the most studied forms considered in the literature:
\begin{equation}
\Upsilon=
\left\{
\begin{array}{l}
C_\Upsilon \frac{T^3}{\phi^2}, \\ C_\Upsilon T, \\ C_\Upsilon \frac{T^3}{M_{\rm Pl}^2},
\end{array}
\right.
\label{Upsilon}
\end{equation}
where in each case, $C_\Upsilon$ is a dimensionless constant.
The model with $\Upsilon \propto T^3/\phi^2$ is motivated by supesymmetry (SUSY) realizations of WI~\cite{Berera:2008ar}.
An investigation of the observational consequences of this dissipation coefficient was performed  e.g. in
ref.~\cite{Bartrum:2013fia}. The model with $\Upsilon \propto T$ was realized for the first time in ref.~\cite{Bastero-Gil:2016qru}.
It is also a limiting case of axion WI when in the presence of massive fermions~\cite{Berghaus:2020ekh,Drewes:2023khq,Berghaus:2024zfg}.
{}Finally, the case $\Upsilon \propto T^3$ is realized in axion-like theories with pure Yang-Mills gauge field. 
It originates as a consequence of sphaleron decay in a thermal bath~\cite{Berghaus:2019whh,Laine:2021ego}.

\section{Examples of results} 
\label{sec6}

\begin{center}
\begin{figure}[!bth]
\centerline{\includegraphics[width=7.cm]{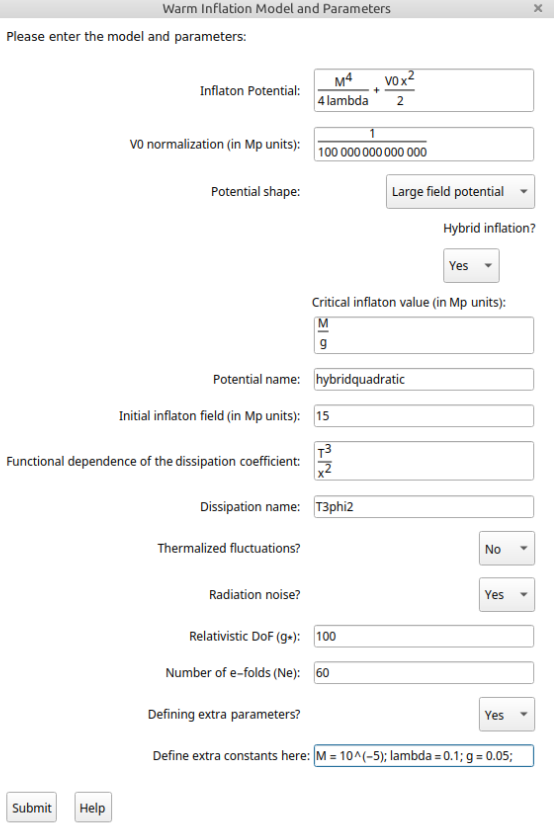}}
\caption{An example of inputs to the fields available in the \texttt{ ParametersWI[]} interface.}
\label{fig1}
\end{figure}
\end{center}

The first step when using the code is to load the \texttt{WI2easy} package as described in section~\ref{sec4}.
Then one runs the comand \texttt{ ParametersWI[]}, which will set all the basic parameters.
In figure~\ref{fig1}, we illustrate its use when considering the hybrid potential with
the dissipation coefficient $\Upsilon \propto T^3/\phi^2$. 

\subsection{The function $G(Q)$} 

The next steps are generating the initial conditions for the chosen model and number of {\it e}-folds through the
command \texttt{FindICs[Qi,Qf]} and then finding the data points for the function $G(Q)$ through  \texttt{FindGQ[]}.
In most of our examples, we set the values of $Q_i$ and $Q_f$ to be $Q_i=10^{-9}$ and $Q_f=4\times 10^3$, respectively.
The procedure can be repeated to all other potentials and dissipation coefficients.
{}For all the results shown, unless otherwise explicitly specified, we have set $N_e=60$ for the total
number of {\it e}-folds of inflation,  $g_*=100$ and $V_0/M_{\rm Pl}^4=10^{-14}$.

\begin{center}
\begin{figure*}[!bth]
\subfigure[$C_\Upsilon T^3/\phi^2$]{\includegraphics[width=7.5cm]{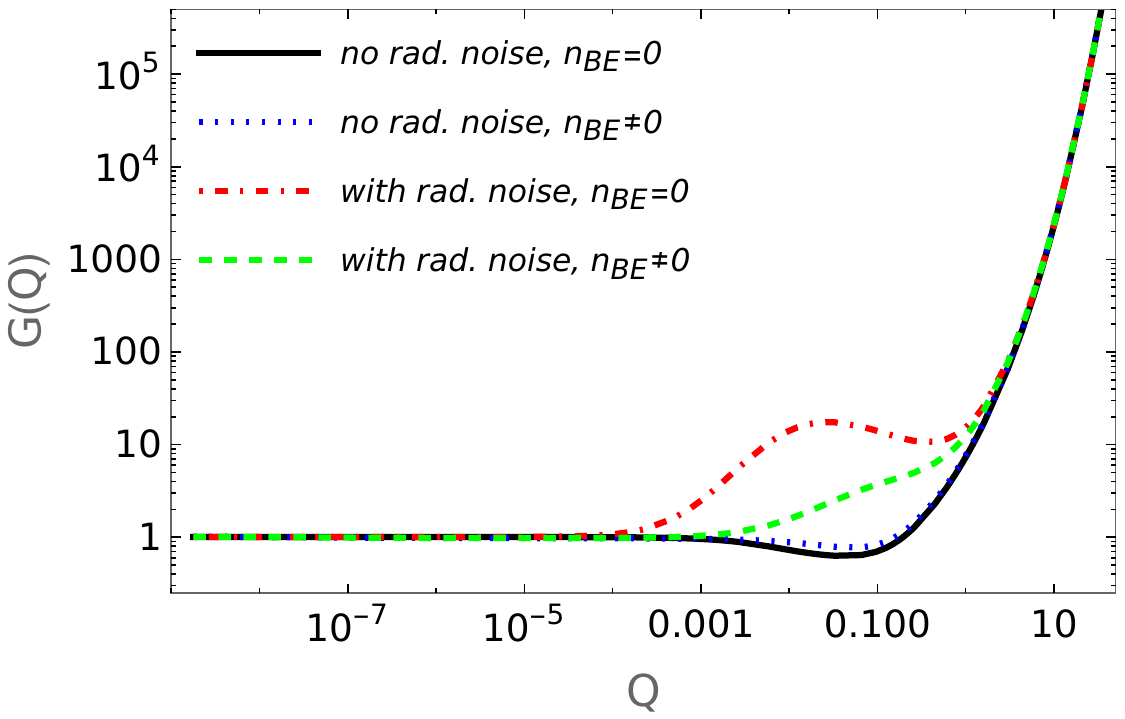}}
\subfigure[$C_\Upsilon T$]{\includegraphics[width=7.5cm]{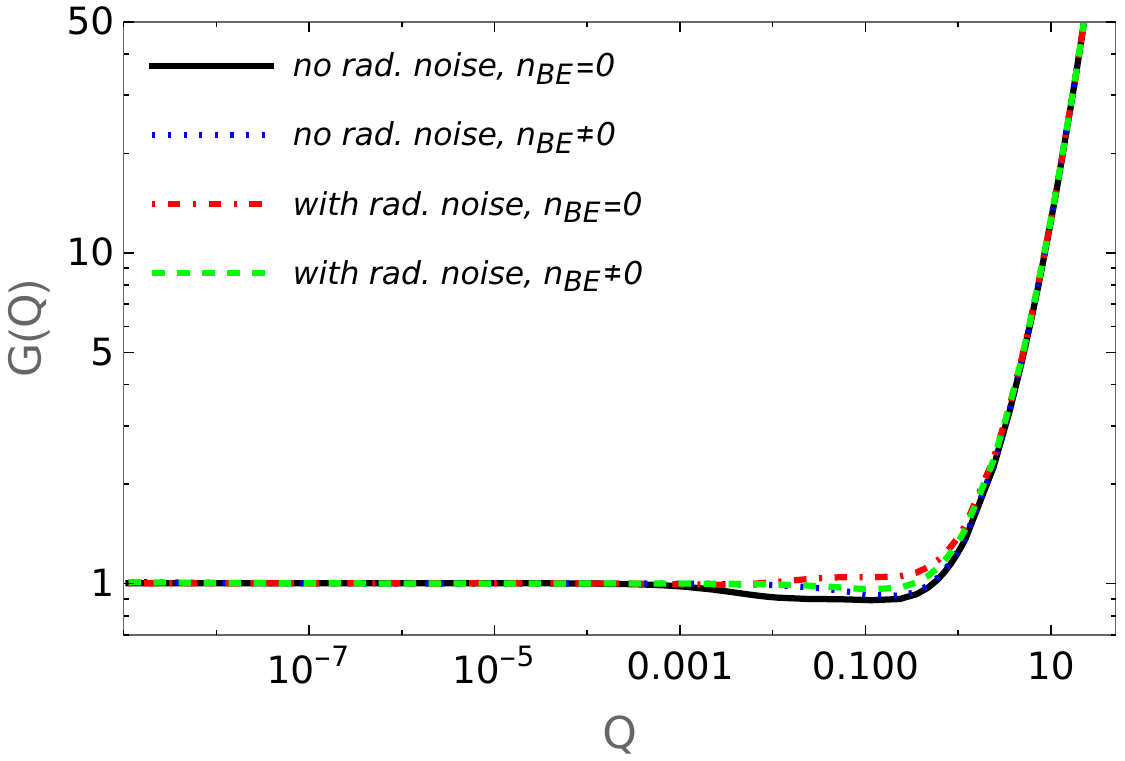}}
\centerline{\subfigure[$C_\Upsilon T^3/M_{\rm Pl}^2$]{\includegraphics[width=7.5cm]{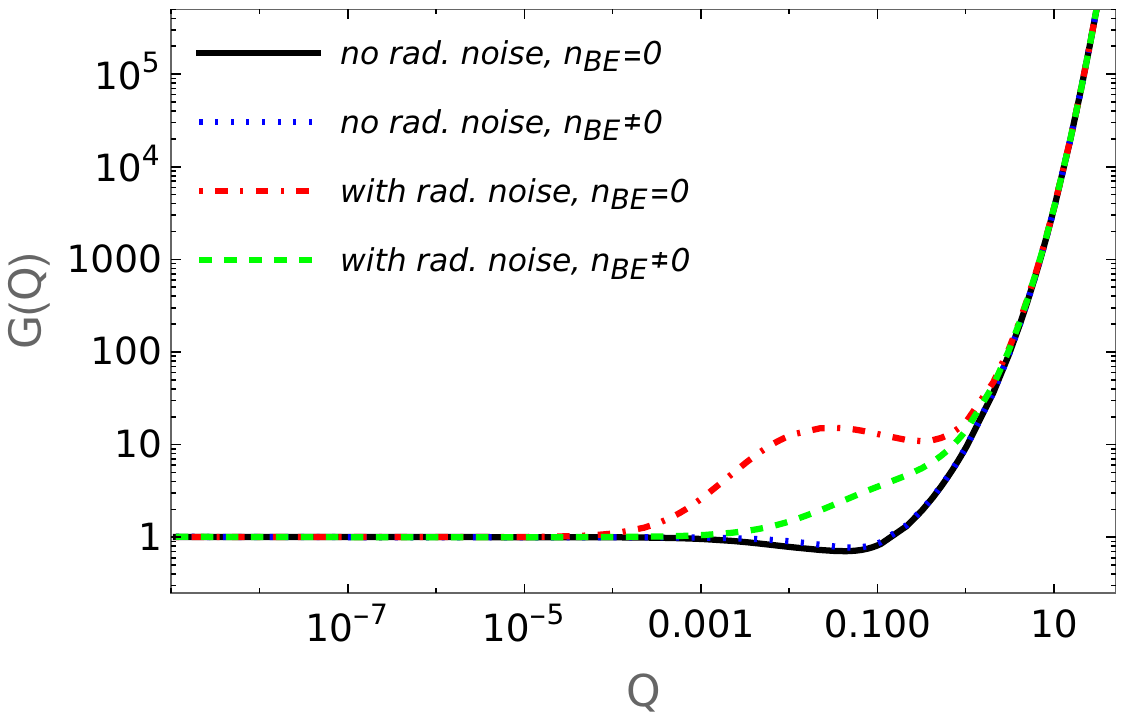}}}
\caption{Results for $G(Q)$ for a quartic monomial potential and different choices
of dissipation coefficients and parameters.}
\label{fig2}
\end{figure*}
\end{center}

In the figure~\ref{fig2} we compare the differences in the function $G(Q)$ when considering the cases of
non-thermal and thermal inflaton fluctuations and also by adding or not the radiation noise term (see explanations
in sec.~\ref{sec3}) for the case of the quartic monomial potential eq.~(\ref{phi4}).

{}From the results of fig.~\ref{fig2}, we can see that the choices of thermal or northermal inflaton fluctuations
and whether the radiation noise term is included or not affect most the region of $Q$ values in the range
$Q \in (10^{-4},10)$. There are no appreciable differences in $G(Q)$ either in the very weak dissipation regime,
$Q\lesssim 10^{-4}$, or in the very strong dissipation regime, $Q \gtrsim 10$. By also comparing panels (a) and
(c) in  fig.~\ref{fig2}, we see that the inflaton dependence in the dissipation coefficient affects
little the behavior of $G(Q)$, being it mostly affected by the temperature dependence. This was also
first seen in ref.~\cite{Bastero-Gil:2011rva} and also recently shown in ref.~\cite{Montefalcone:2023pvh}.

\begin{center}
\begin{figure*}[!bth]
\subfigure[with radiation noise]{\includegraphics[width=7.5cm]{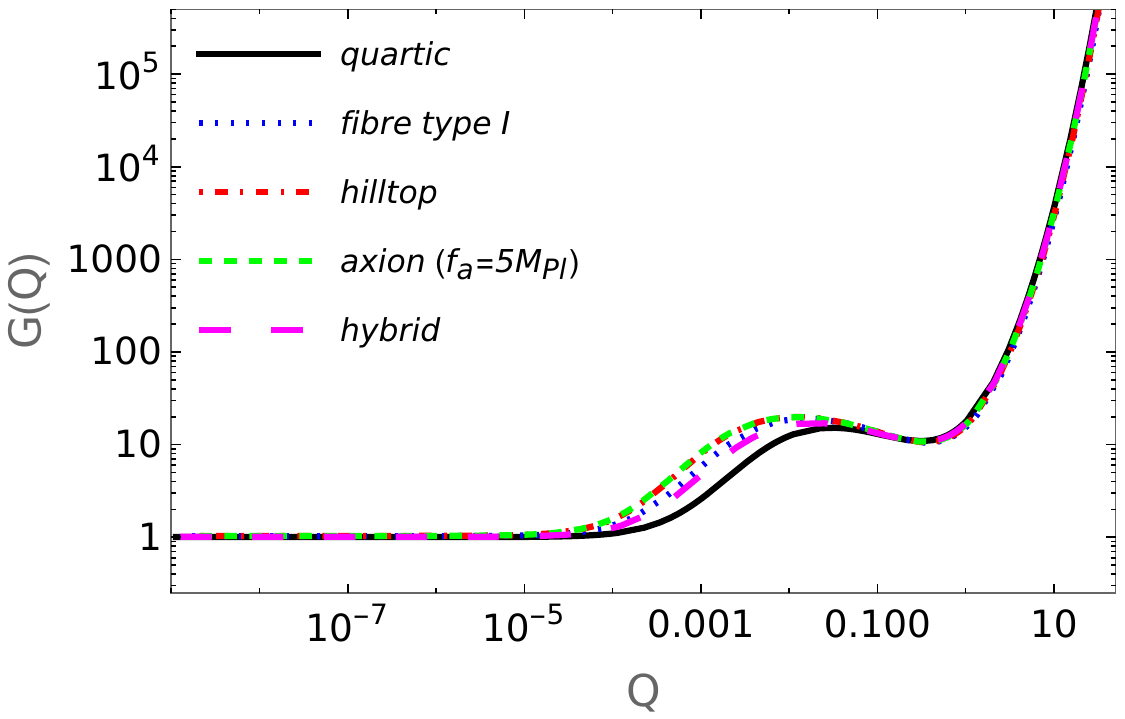}}
\subfigure[no radiation noise]{\includegraphics[width=7.5cm]{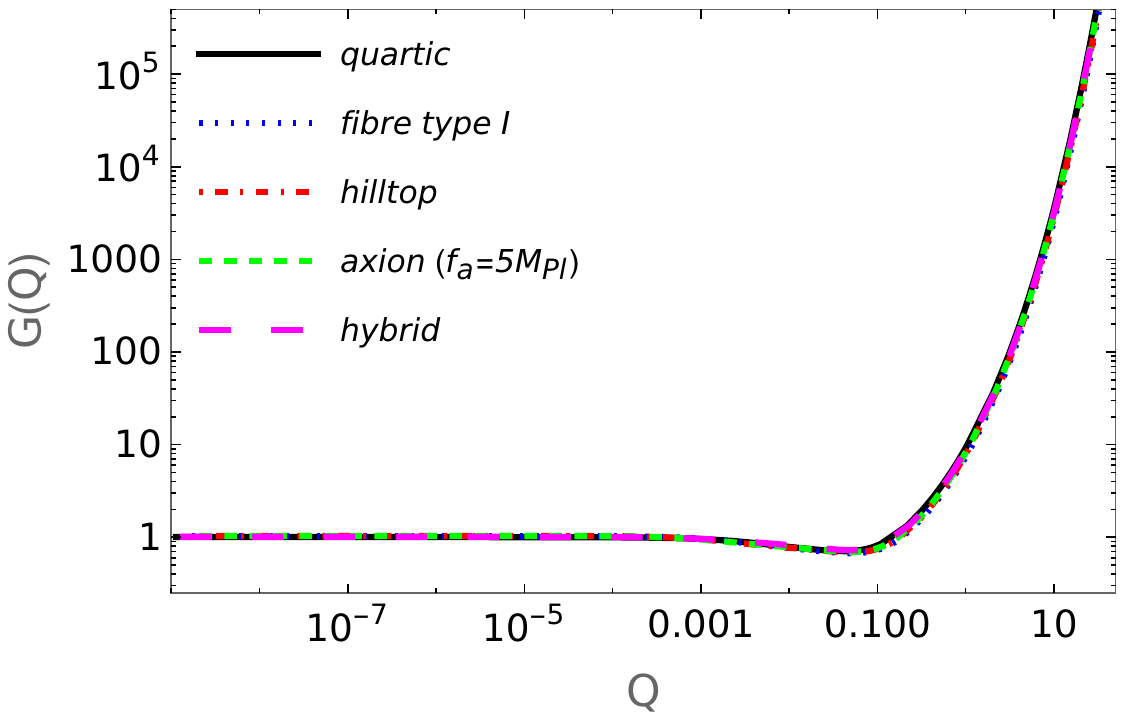}}
\caption{Results for $G(Q)$ for the different forms of potentials considered
here, when including (panel a) or not including (panel b) the radiation noise term.
In both cases we are considering non thermal inflaton perturbations and for a dissipation
coefficient $\Upsilon = C_\Upsilon T^3/M_{\rm Pl}^2$. }
\label{fig3}
\end{figure*}
\end{center}

\begin{center}
\begin{figure*}[!bth]
\subfigure[with radiation noise]{\includegraphics[width=7.5cm]{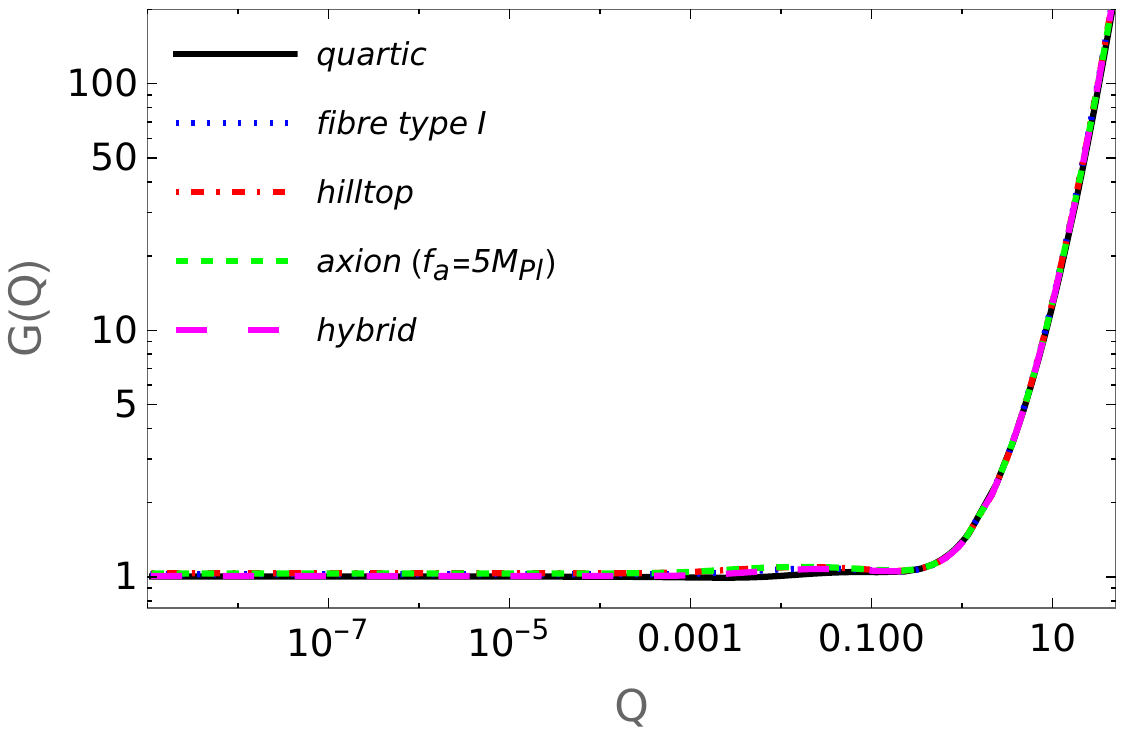}}
\subfigure[no radiation noise]{\includegraphics[width=7.5cm]{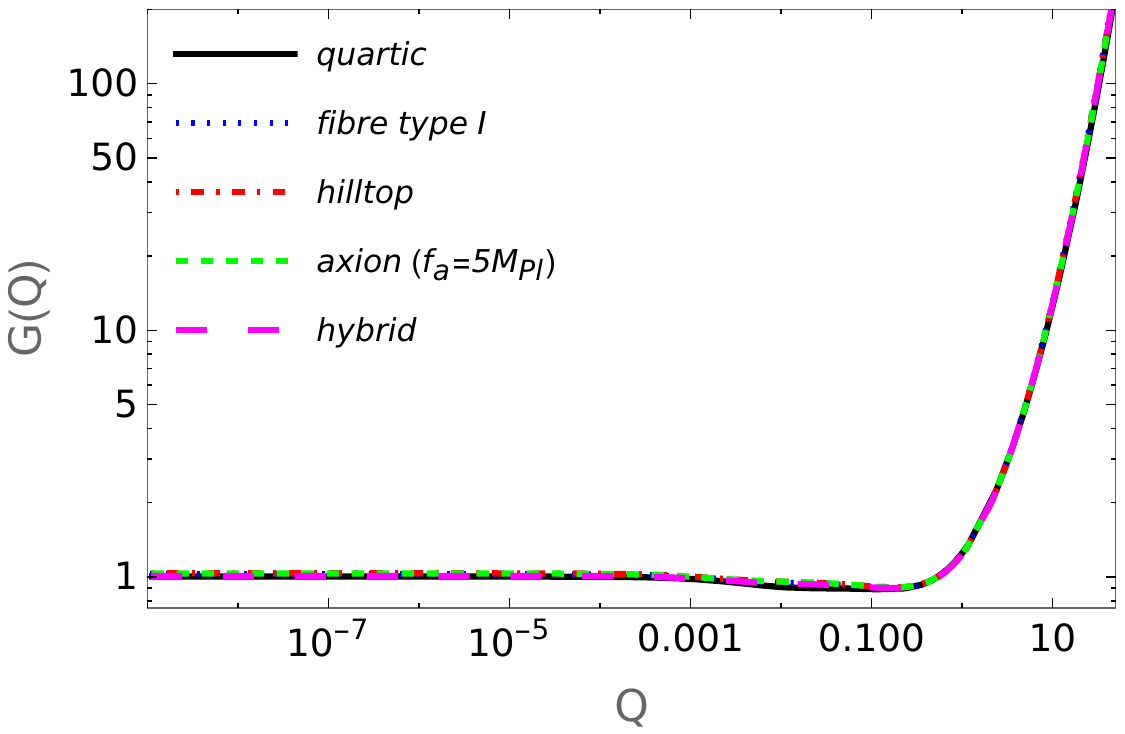}}
\caption{Same as in fig.~\ref{fig3}, but now for a dissipation
coefficient $\Upsilon = C_\Upsilon T$. }
\label{fig4}
\end{figure*}
\end{center}

Previous work~\cite{Montefalcone:2023pvh} demonstrated a universality in the functional form of $G(Q)$: for a fixed dissipation coefficient $\Upsilon$, distinct inflaton potentials yield approximately identical $G(Q)$ profiles. This universality implies $G(Q)$ depends primarily on  $\Upsilon$, not $V(\phi)$, under conditions of neglecting radiation noise and assuming thermalized inflaton perturbations (as in ref.~\cite{Montefalcone:2023pvh}). Here, we investigate whether this universality persists when incorporating radiation noise in the perturbation equations and relaxing thermalization assumptions (figs.~\ref{fig3} and \ref{fig4}). We find that without the radiation noise term, 
the universality holds even for nonthermal inflaton perturbations. However, with the radiation noise term included, the universality breaks down, as seen most pronounced from the results of fig.~\ref{fig3}(a). {}For $Q\in (10^{-4},10)$, $G(Q)$ exhibits marked potential-dependent variations (as also seen in fig.~\ref{fig2}), which are amplified for temperature-sensitive dissipation coefficients (e.g., for $\Upsilon \propto T^c$, with $c>1$).
These results underscore the critical role of radiation noise and thermalization in shaping $G(Q)$, necessitating first-principles computations for precision cosmology in WI.

\subsection{Comparison with previous results for $G(Q)$} 

Let us here compare our results for $G(Q)$ with existing ones that are commonly
used in the WI literature. The earliest proposed forms of $G(Q)$ were fitting
functions of the numerical data obtained with the numerical solutions of the 
stochastic equations for the perturbations in WI, first given in ref.~\cite{Graham:2009bf} 
and then in ref.~\cite{Bastero-Gil:2011rva}. These fitting functions were then improved
in later publications and with different degrees of 
complications~\cite{Benetti:2016jhf,Motaharfar:2018zyb,Das:2020xmh}. The most recent
ones are from ref.~\cite{Montefalcone:2023pvh}. All these previous results for $G(Q)$ were
obtained in the case of thermal inflaton perturbations, thus taking $n_* \equiv n_{\rm BE}$
in eq.~(\ref{PR1}), and also neglecting the radiation noise term 
in eq.~(\ref{deltadotrhor})\footnote{One exception to these is the fitting proposed
in ref.~\cite{Mirbabayi:2022cbt}, which was produced for a dissipation coefficient
$\Upsilon \propto T^3$ for the cases of nonthermal ($n_*=0$) and in the presence
of the radiation noise term in eq.~(\ref{deltadotrhor}). This case is analyzed in details
in a dedicate separate publication~\cite{ORamos:2025uqs}.}. Hence, we compare the present results under
the same conditions. 

\begin{center}
\begin{figure}[!bth]
\centerline{\includegraphics[width=8.2cm]{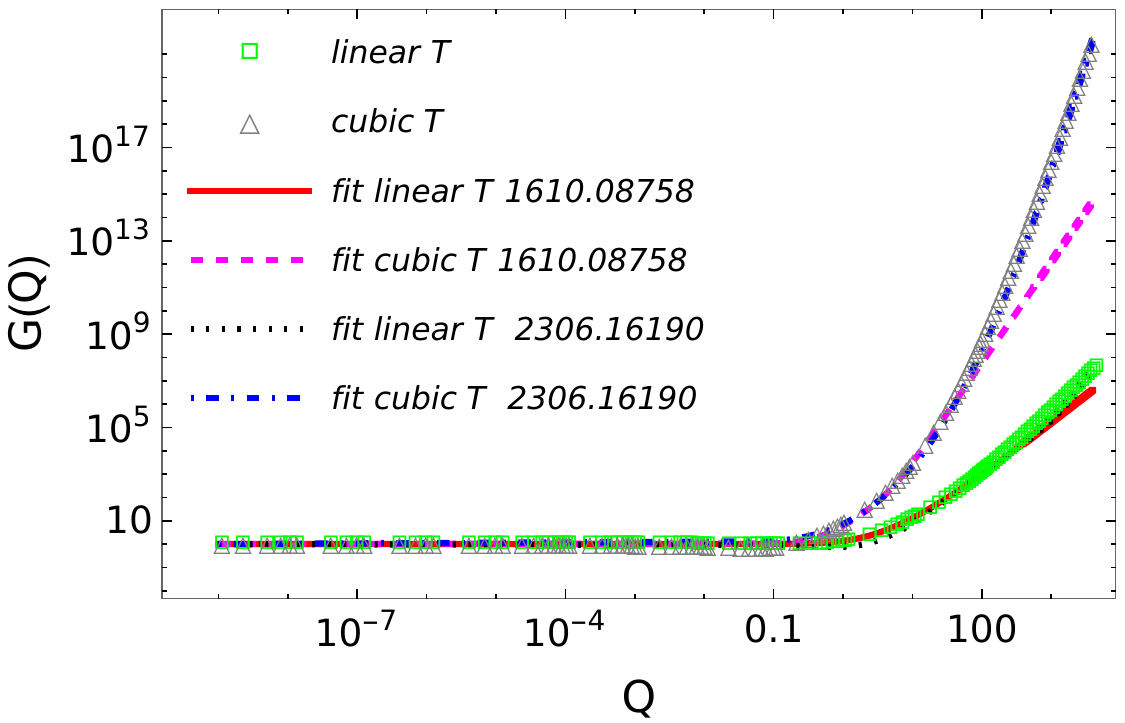}}
\caption{Comparison of the data for $G(Q)$, obtained with this code for the cases
of the dissipation coefficients $\Upsilon \propto T $ and $\propto T^3$,
with previous fitting formulas from refs.~\cite{Benetti:2016jhf} and \cite{Montefalcone:2023pvh}.}
\label{fig5}
\end{figure}
\end{center}

In {}Fig.~\ref{fig5}, we compare our results with earlier parameterizations of $G(Q)$ 
from the literature, focusing on the two most widely used forms for dissipation coefficients
$\Upsilon \propto T $ and $\propto T^3$
(see, e.g. refs.~\cite{Benetti:2016jhf,Motaharfar:2018zyb}) as well as recent fits derived from 
\texttt{WarmSPy}~\cite{Montefalcone:2023pvh}.
The earlier fitting functions tend to under estimate\footnote{We note that the earlier
fitting formulas for $G(Q)$ were designed to better fit the power spectrum in the
low $Q$ region. The reason for this was that the existing models of WI at the time
were only able to produce results consistent with the observations in the weak
dissipative regime~\cite{Benetti:2016jhf}.}  $G(Q)$ for $Q \gtrsim 50$.
The most recent fittings from ref.~\cite{Montefalcone:2023pvh} 
accommodate recent WI scenarios where strong dissipation enhances observational signatures 
(e.g., suppressed tensor modes). Our analysis highlights the necessity of updated $G(Q)$ parameterizations 
for precision cosmology in high-$Q$ regimes, particularly for temperature-sensitive dissipation coefficients.

As already commented before, here we do not offer any new fitting formulas for $G(Q)$. 
\texttt{WI2easy}, through the command  \texttt{FindGQ[]}, generates sufficient data points
from which an internal function $G(Q)$ is created through a spline interpolation.
This is then used in the subsequent parts of the code, e.g. in the modules
\texttt{ObservationsWI} and  \texttt{QrangeWI}. We find this approach to be more efficient and
to lead to better precision when obtaining the perturbation quantities that can be
direct compared with the observation data. 

\subsection{Background quantities and the power spectrum} 

Having obtained the data points for $G(Q)$, the code is now ready to properly normalize
the amplitude of the scalar of curvature power spectrum and to obtain the evolution of
background and perturbation quantities. This is achieved by running the command
\texttt{ObservationsWI[Q0value,x0value]}. 

In the example notebook \texttt{main.nb}, we have considered the quartic inflaton potential
eq.~(\ref{phi4}) with a dissipation coefficient that is linear in the temperature. This model was first
studied in a viable quantum field theory model for WI in ref.~\cite{Bastero-Gil:2016qru}.
In that reference, it was considered both nonthermal and thermal inflaton fluctuations,
but neglecting the radiation noise term discussed in sec.~\ref{sec3} and using fixed values
of {\it e}-folds ($50$ and $60$) for Hubble radius crossing. Let us obtain here the case
of thermal inflaton perturbations and by including the radiation noise term in the radiation
perturbation equation. {}Furthermore, as already explained, the code automatically computes
the appropriate $N_*$ value where Hubble radius crossing happens and with the proper
normalization $V_0$ for the potential. 

In the example notebook, we have considered the value $Q_*=1$, for which we obtain
$V_0/M_{\rm Pl}^4\simeq 3.20 \times 10^{-15}$ for the normalization of the potential, 
$C_\Upsilon \simeq 0.025$ for the dimensionless constant in the dissipation coefficient,
and perturbation quantities $r\simeq 3.37 \times 10^{-4}$ for the tensor-to-scalar ratio,
$n_s\simeq 0.9688$ for the spectral tilt and $\alpha_s \simeq -2.85 \times 10^{-4}$ for the
running, with $N_* \simeq 58.2$.

\begin{center}
\begin{figure*}[!bth]
\subfigure[]{\includegraphics[width=5.cm]{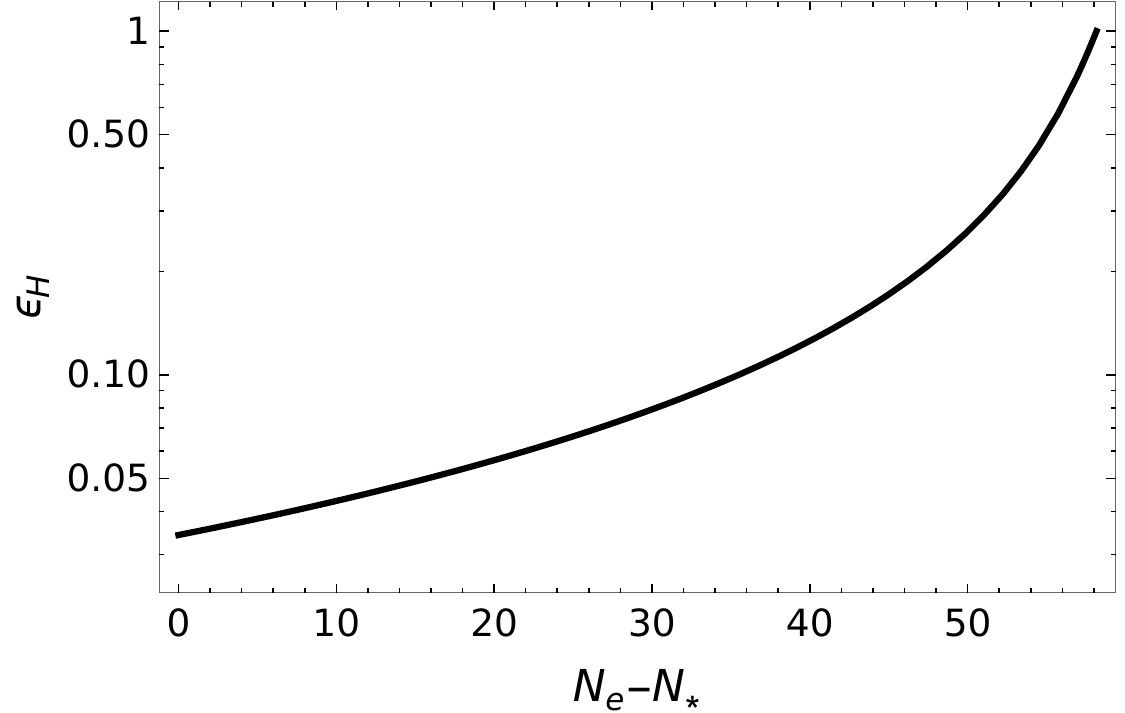}}
\subfigure[]{\includegraphics[width=5.cm]{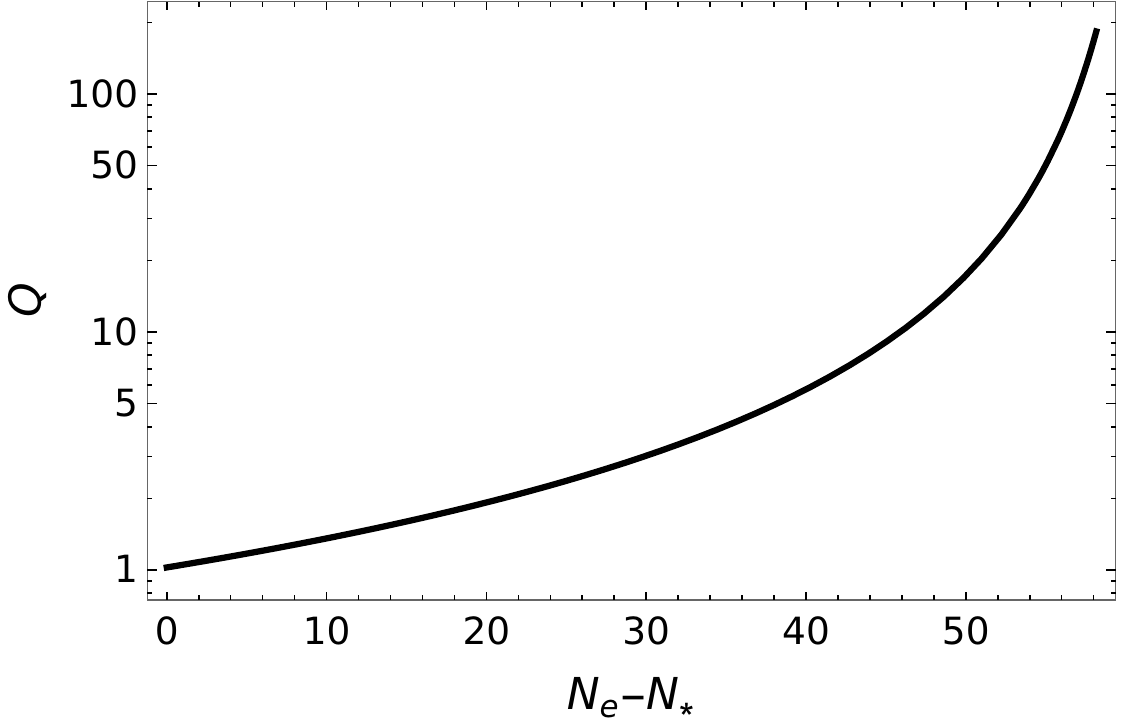}}
\subfigure[]{\includegraphics[width=5.cm]{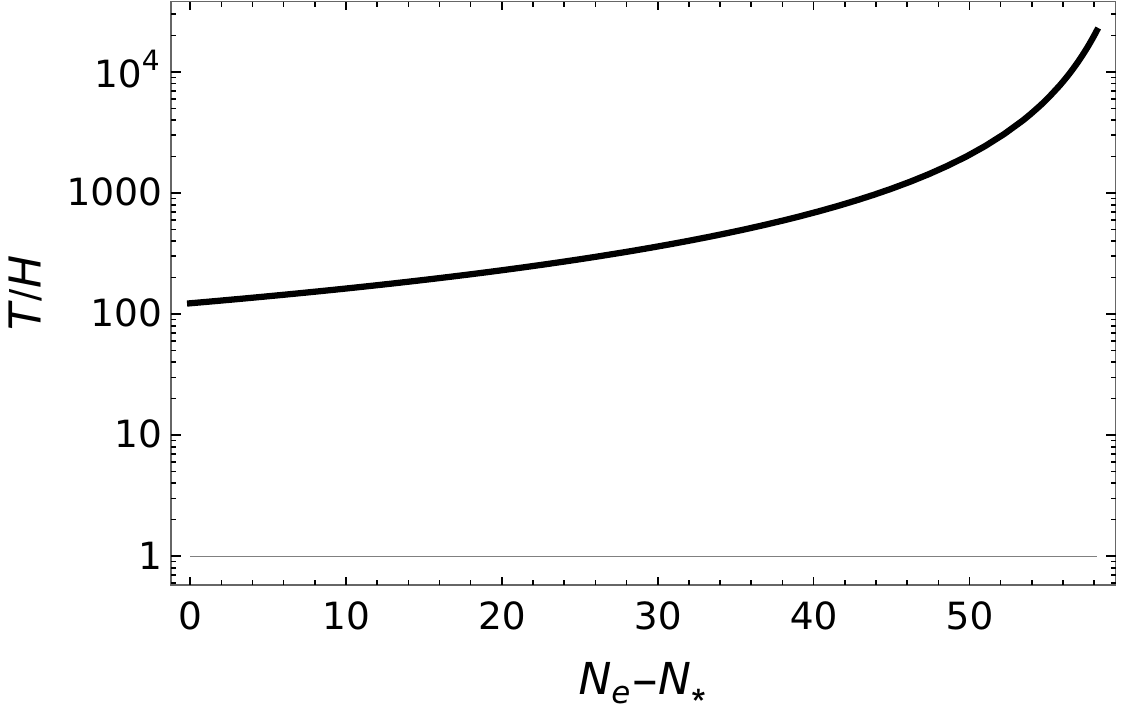}}
\subfigure[]{\includegraphics[width=5.cm]{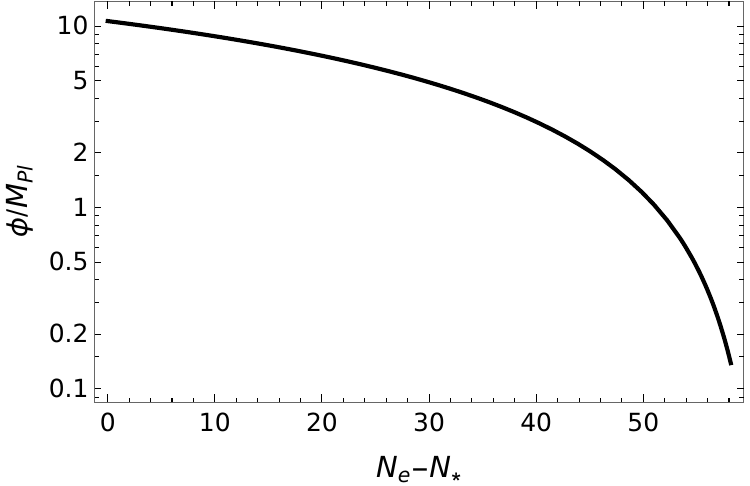}}
\subfigure[]{\includegraphics[width=5.cm]{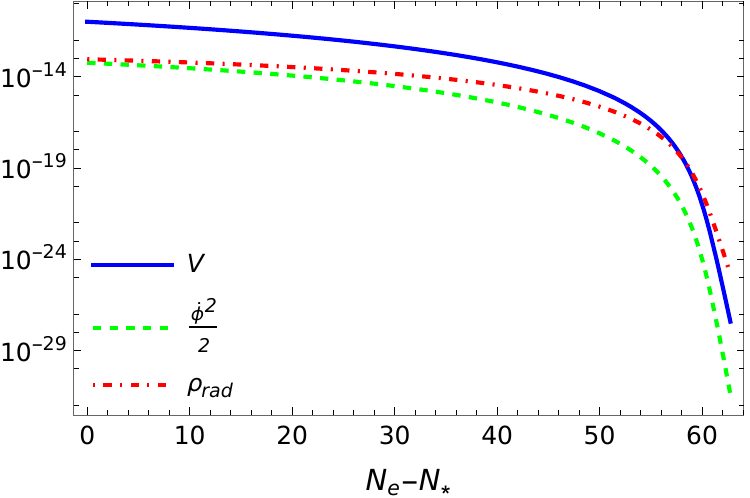}}
\subfigure[]{\includegraphics[width=5.cm]{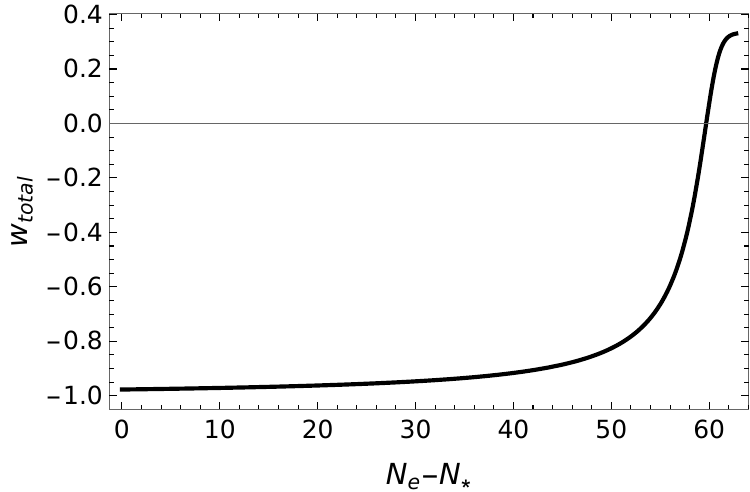}}
\caption{Evolution of some of the background quantities available through
the command \texttt{MakePlotsEvolution}.}
\label{fig6}
\end{figure*}
\end{center}

Still in the module \texttt{ObservationsWI[Q0value,x0value]}, the use of the command \break
 \texttt{MakePlotsEvolution}
automatically generates plots for the evolution of several background quantities for the chosen
value of $Q_*$.  
In fig.~\ref{fig6} we show the evolution for $\epsilon_H$, $Q$, $T/H$, $\phi$, the energy
densities ($V(\phi)$, $\dot\phi^2/2$ and $\rho_r$, all in units of the reduced Planck mass)
and the total equation of state $w_{\rm total}$ for the example we are considering here.

\begin{center}
\begin{figure*}[!bth]
\subfigure[]{\includegraphics[width=5.cm]{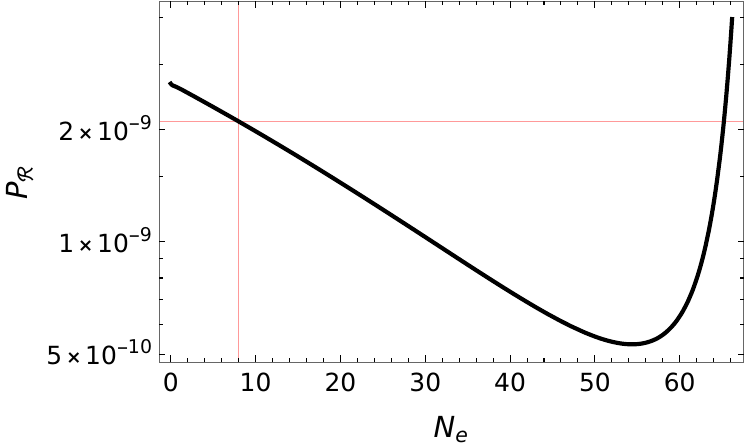}}
\subfigure[]{\includegraphics[width=5.cm]{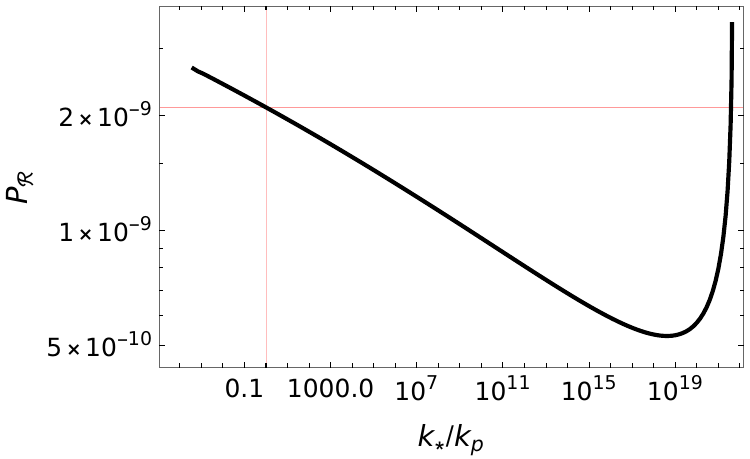}}
\subfigure[]{\includegraphics[width=5.cm]{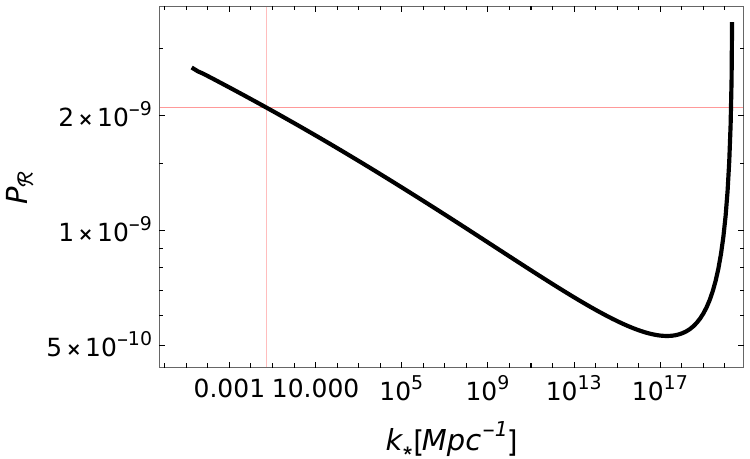}}
\caption{The scalar of curvature power spectrum obtained with
the command \texttt{MakePlotsEvolution} for the value of $Q_*$ chosen.}
\label{fig7}
\end{figure*}
\end{center}

Through the same command, plots for the scalar of curvature power spectrum is also produced.
This is illustrated in fig.~\ref{fig7}. These are obtained directly from the proxy expression
eq.~(\ref{PR1}) corrected by the function $G(Q)$ from eq.~(\ref{powersG}). 
Note that we can always conveniently relate a comoving scale $k_*$ at
Hubble-exit epoch through $k_*=aH$. In the plots shown in fig.~\ref{fig7}, the thin
vertical line indicates the pivot scale $k_p$ (which in the code it
is given by the  default value $k_p=0.05 {\rm Mpc}^{-1}$), while the horizontal thin line
is at the value for the CMB amplitude normalization (with default value $A_s \simeq 2.105 \times 10^{-9}$ 
as considered in the code).

\subsection{Dependence of the background and perturbation quantities on the value of $Q$} 

Our final command is \texttt{QrangeWI[Qinit]}. As explained earlier, it repeats the command
\texttt{ObservationsWI} but now for different values of $Q_*$ and save the data for the
background and perturbation quantities as a function of $Q_*$.
Still considering the same example of a quartic monomial potential with a linear in the
temperature dissipation coefficient, we give the results obtained through the example
notebook when using this command.  

\begin{center}
\begin{figure*}[!bth]
\subfigure[]{\includegraphics[width=5.cm]{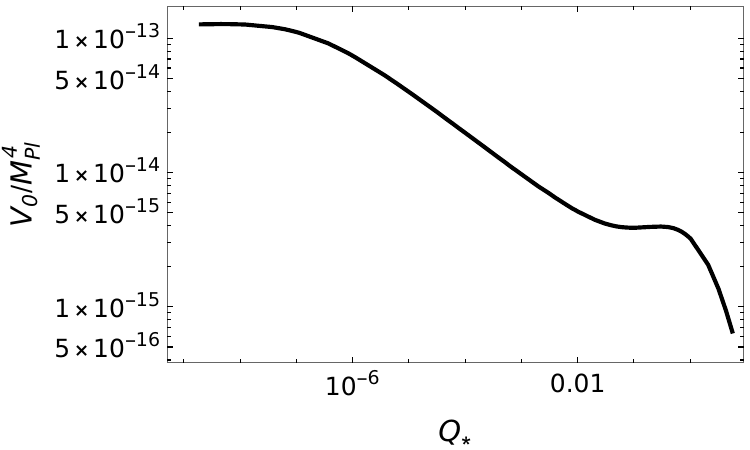}}
\subfigure[]{\includegraphics[width=5.cm]{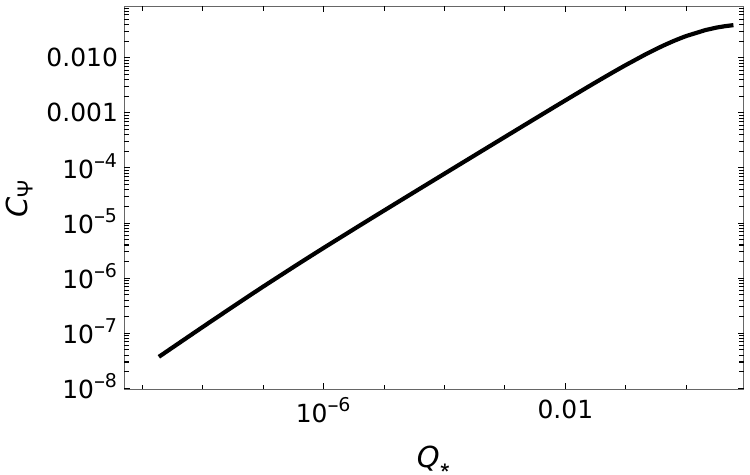}}
\subfigure[]{\includegraphics[width=5.cm]{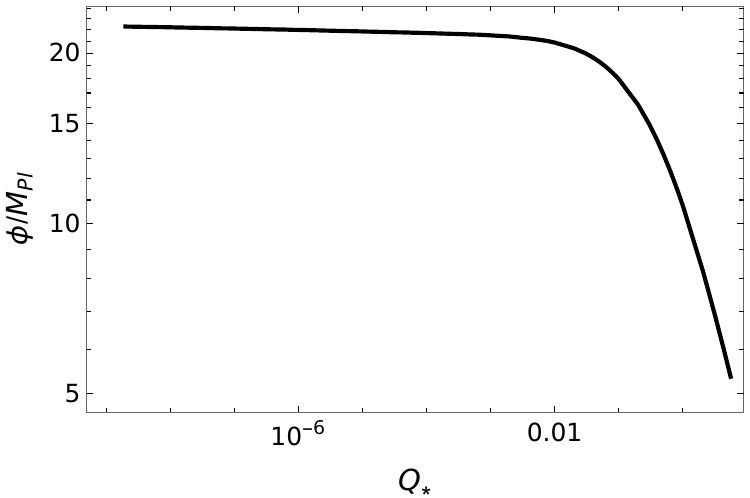}}
\subfigure[]{\includegraphics[width=5.cm]{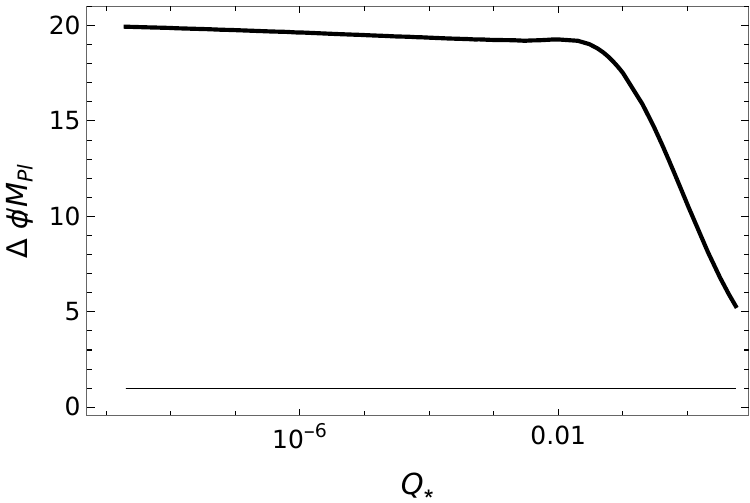}}
\subfigure[]{\includegraphics[width=5.cm]{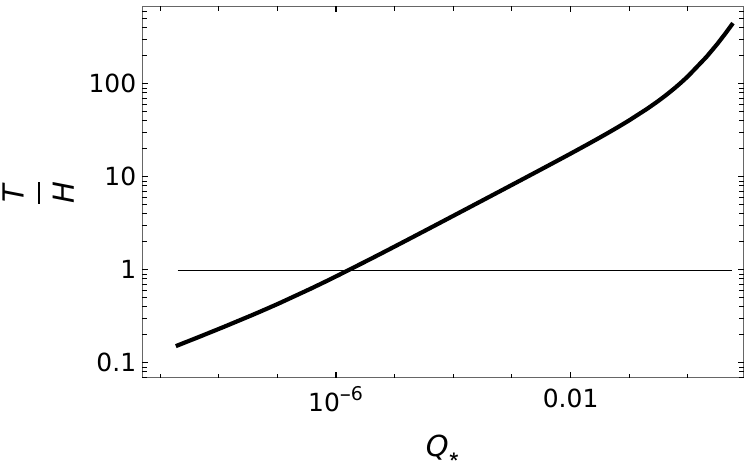}}
\subfigure[]{\includegraphics[width=5.cm]{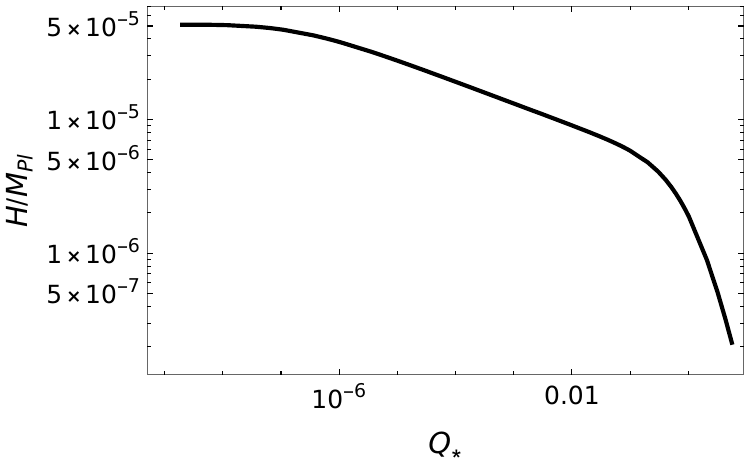}}
\subfigure[]{\includegraphics[width=5.cm]{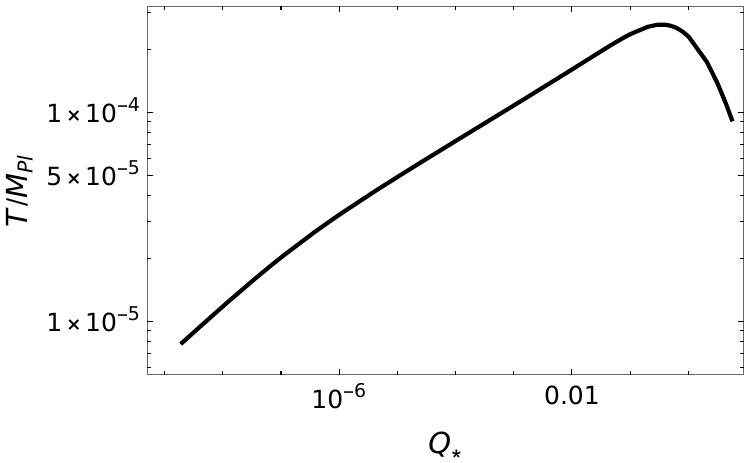}}
\subfigure[]{\includegraphics[width=5.cm]{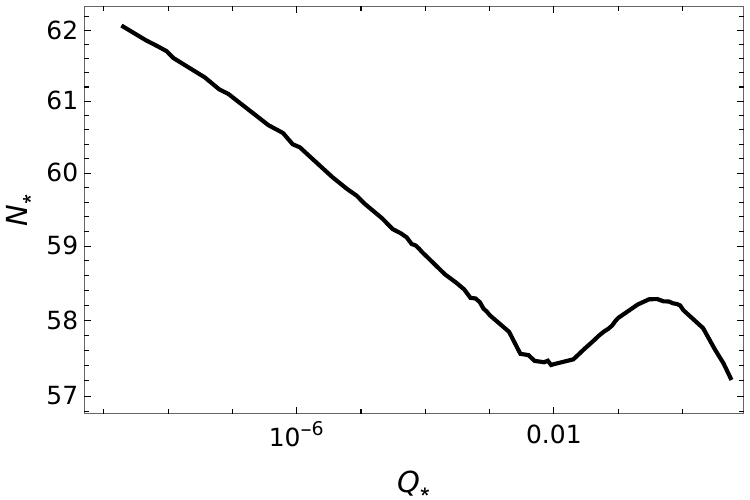}}
\caption{Dependence of the background quantities as a function of $Q_*$ (obtained
at the Hubble-exit point  $N_*$).}
\label{fig8}
\end{figure*}
\end{center}

In fig.~\ref{fig8} we show some of the results that are obtained for the background
quantities as a function of $Q_*$.
{}From panels (a) to (h), we have the potential normalization $V_0$, the dissipation
coefficient $C_\Upsilon$, the inflaton amplitude, the variation of the inflaton amplitude
in between the Hubble-exit ($N_*$ {\it e}-folds before the end of inflation)
and at the end of inflation,  the ratio $T/H$, the value of the Hubble parameter,
the temperature and the  Hubble-exit point ($N_*$) all as a function of $Q_*$.

{}From the fig.~\ref{fig8}(d) we can see that the inflaton field variation tends to move
towards sub-Planckian values, which is important from an effective field theory 
viewpoint~\cite{Motaharfar:2018zyb}. {}From fig.~\ref{fig8}(e), we can see that 
WI effectively starts for $Q \gtrsim 10^{-6}$. This is a characteristic that we have
also seen when testing other types of potentials and dissipation functions in WI.  
We also note from fig.~\ref{fig8}(g) that towards large values
of $Q$ the temperatures reaches a peak and then starts to drop. This behavior is
closely related to the reason of why the power spectrum seen in fig.~\ref{fig7} displays
more power towards the end of inflation (smaller scales). This effect 
has been explored recently in connection to primordial black hole formation and
magnification of gravitational waves in WI~~\cite{Arya:2019wck,Bastero-Gil:2021fac,Correa:2022ngq,Ferraz:2024bvd,Ito:2025lcg}.

\begin{center}
\begin{figure*}[!bth]
\subfigure[]{\includegraphics[width=7.5cm]{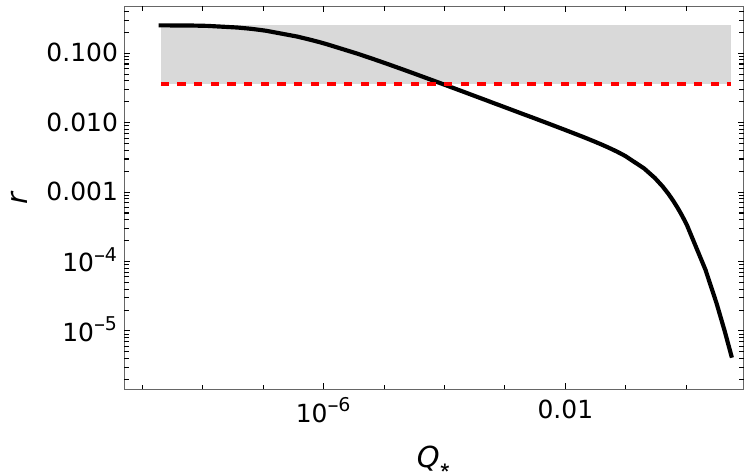}}
\subfigure[]{\includegraphics[width=7.5cm]{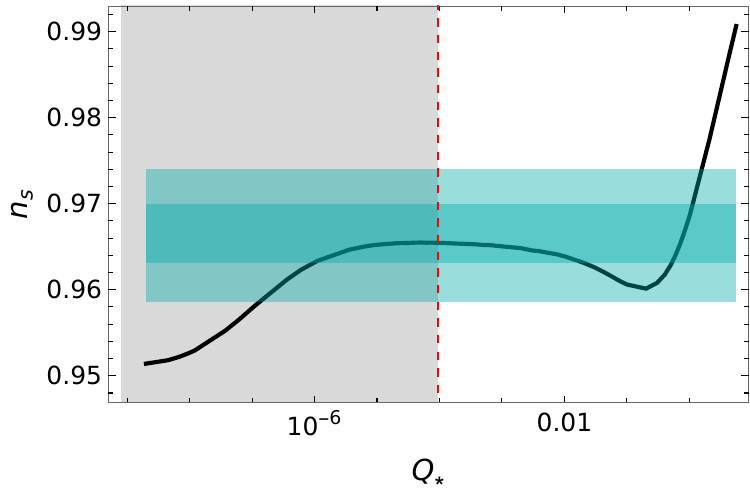}}
\subfigure[]{\includegraphics[width=7.5cm]{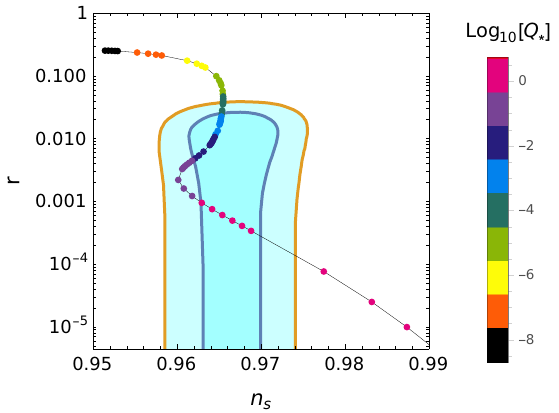}}
\subfigure[]{\includegraphics[width=7.5cm]{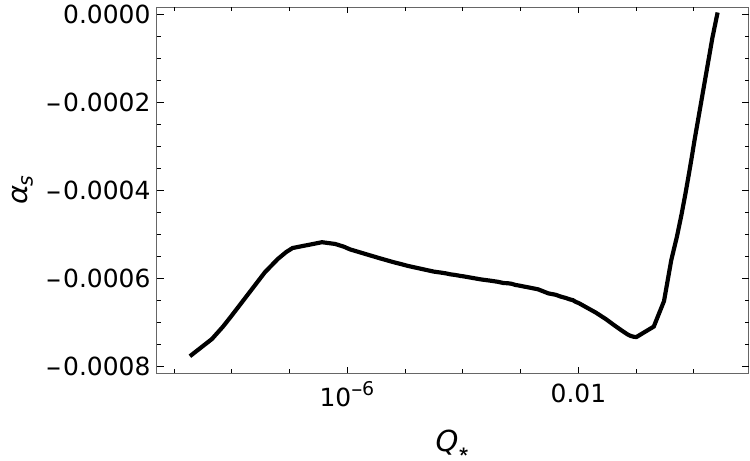}}
\caption{Perturbation quantities when varying $Q_*$. The top gray region in panel (a)
and the left gray region in panel (b) indicate the region where $r> 0.036$, with the upper bound on $r$ from the
BICEP, Keck Array and Planck combined data~\cite{BICEP:2021xfz}. The cyan regions in panels (b)
and (c) indicate the one- and two-sigma constraints, also from ref.~\cite{BICEP:2021xfz}. }
\label{fig9}
\end{figure*}
\end{center}

{}Finally, in fig.~\ref{fig9} we give the perturbation quantities obtained for the
quartic monomial potential with a linear in the
temperature dissipation coefficient (and considering thermal inflaton perturbations
and the radiation noise term in the radiation perturbation equation)\footnote{This result also
compares well with the one obtained in \cite{Ballesteros:2023dno}, shown by the purple line 
in fig. 11 of that reference, except that here
we have chosen $g_* = 100$, while in that reference the authors 
considered $g_* = 12$, as motivated from \cite{Bastero-Gil:2016qru}.}.
The cyan regions in the figure are the one- and two-sigma constraints\footnote{Note that
recently, the Atacama Cosmology Telescope (ACT) have reported new constraints which
shifts the spectral tilt towards slight larger values. In WI, and in special in the example
model studied here, this implies that slight larger values of $Q_*$ are allowed, increasing the
viability of WI towards the strong dissipation regime~\cite{Berera:2025vsu}.}
obtained from the BICEP, Keck Array and Planck combined data~\cite{BICEP:2021xfz}.

\section{Conclusion} 
\label{conclusions}

In this paper, we introduce \texttt{WI2easy}, a versatile \texttt{Mathematica} code designed for analyzing the 
dynamics of Warm Inflation (WI). This comprehensive tool supports a broad range of inflaton potentials--including 
large-field, small-field, and hybrid models--enabling researchers to explore diverse WI scenarios. The code also 
accommodates generic dissipation coefficients, which can be arbitrary functions of temperature and inflaton field amplitude. 
With its adaptable framework, \texttt{WI2easy} streamlines numerical investigations of WI phenomenology while 
maintaining computational efficiency and user accessibility.

The code advances the analysis of WI perturbations by employing a {}Fokker-Planck formalism
(originally formulated in \cite{Ballesteros:2022hjk,Ballesteros:2023dno}), 
bypassing the direct numerical solution of stochastic differential equations. This approach 
transforms the computation of the primordial power spectrum--a statistical quantity derived 
from two-point correlation functions--into the deterministic evolution of a coupled matrix 
differential equation system, integrated alongside the background equations. Compared to 
traditional stochastic methods for WI perturbations, this framework offers significant 
computational advantages: (a) there is a gain in speed -- the deterministic system 
eliminates the need for ensemble averaging over stochastic realizations and (b) 
precision -- leveraging \texttt{Wolfram Mathematica}'s high-precision numerical solvers, 
the code allows users to tailor integration methods (e.g., adaptive step sizes, 
stiffness handling) to their specific model requirements.
Additionally, the code prioritizes accessibility. Its modular design ensures 
straightforward execution for standard use cases while retaining flexibility--users 
can easily modify default parameters or implement model-specific adjustments without 
advanced programming expertise.

The code offers broad applicability, generating diverse datasets critical for 
advancing WI research. A key output is the power spectrum as a function of the 
comoving scale, which facilitates statistical analysis of WI models through 
seamless integration with CMB tools like CAMB. This enables rigorous testing of WI 
scenarios against both current and forthcoming observational datasets. Earlier 
studies--from pioneering works~\cite{Benetti:2016jhf,Bastero-Gil:2017wwl,Arya:2017zlb} 
to recent investigations~\cite{Kumar:2024hju,Santos:2024pix,Santos:2024plu}--have 
relied on simplified parametrizations of the $G(Q)$ function to improve the approximate 
expression for the WI power spectrum (e.g., eq. (\ref{PR1})). In contrast, our code 
provides direct access to the power spectrum for any dissipation coefficient 
$\Upsilon(T,\phi)$ and single-field inflaton potentials $V(\phi)$, eliminating 
reliance on ad hoc approximations. This capability is essential for precision 
comparisons between WI predictions and observational constraints, empowering users 
to systematically explore model viability across the full parameter space of 
dissipative dynamics in the context of WI.

{}Future releases of the code will expand its scope to include multi-field 
inflation models and non-Gaussianity calculations, broadening its applicability 
to more advanced cosmological scenarios. Planned enhancements will also focus on 
computational efficiency: parallelization and code optimizations will accelerate 
parameter space exploration while maintaining high numerical precision. These 
upgrades aim to provide users with faster, more versatile tools for state-of-the-art 
WI analyses, aligning the code's capabilities with evolving theoretical 
and observational demands. 

\appendix
\section{Background and perturbation equations implemented in the code} 
\label{appA}

In the code, the background and perturbation equations enter as a function
of {\it e}-folds instead of time variable.
Using $dN_e = H dt$ and writing the background equations (\ref{phiEOM}) and
\ref{radEOM} as a system of 
first order differential equations, we have that (note however 
that throughout our code, even though we work with the evolution in terms of
{\it e}-folds, we still keep $\dot \phi \equiv y$ as such)
\begin{eqnarray}
\phi'&=& \frac{y}{H},
\label{phiprime}
\\
y' &=& -3(1+Q) y - \frac{V_{,\phi}}{H},
\label{yprime}
\\
\rho_r' &=& -4 \rho_r + 3 Q y^2,
\label{rhorprime}
\end{eqnarray}
with prime denoting derivative with respect to $N_e$.
The above equations together with the differential equation for the 
two-point statistical momenta, ${\bf J}$, eq.~(\ref{Jequation})
then form a complete system.

In eq.~(\ref{Jequation}), the elements for the matrices ${\bf A}$ 
and ${\bf B}$ are read from the perturbation equations  (\ref{deltaddotphi})-(\ref{dotpsir})
and (\ref{metric1})-(\ref{metric3}). {}From the definition of the vector ${\bm \Phi}$,
eq.~(\ref{Phi}), the matrix ${\bf A}$ has elements $a_{ij}$ given by (also remember that
we work in units of reduced Planck mass, which in practice means taking $M_{\rm Pl}=1$):
\begin{eqnarray}
&&a_{11}=0,\; a_{12}=-1,\: a_{13}=0,\;a_{14}=0,
\nonumber \\
&& a_{21}= -\frac{2 y^2 \rho_r}{3 H^4}  -\frac{y^4}{2 H^4}  +\frac{k^2}{(aH)^2} +
\frac{3 y^2}{H^2} + \frac{3 Q y^2}{2 H^2}  + \frac{2 y V_{,\phi}}{ H^3}
 + \frac{V_{,\phi\phi}}{ H^2} + \frac{p y }{H^2},
\nonumber \\
&& a_{22}= 3 + 3Q - \frac{2 \rho_r}{3 H^2}- \frac{y^2}{2 H^2},
\nonumber \\
&& a_{23}= \frac{y}{3 H^3} + \frac{c y T}{4 \rho_r H^2},
\nonumber \\
&& a_{24}= \frac{2 y \rho_r}{3 H^4} + \frac{y^3}{2 H^4} - \frac{3y}{H^2}
- \frac{3 Q y}{2 H^2} - \frac{V_{,\phi}}{H^3},
\nonumber \\
&& a_{31} =  - \frac{\rho_r y^3}{3H^3} + \frac{2 y \rho_r}{H} + \frac{3 Q y^3}{2H}
+ \frac{2 \rho_r V_{,\phi}}{3 H^2}  - \frac{p y^2}{H},
\nonumber \\
&& a_{32}=   \frac{2 y \rho_r }{3 H}  -6 y H Q,
\nonumber \\
&& a_{33} = 4 +  \frac{2 \rho_r}{3H^2}  - \frac{c T y^2}{4 \rho_r H},
\nonumber \\
&& a_{34}= \frac{y^2 \rho_r}{3H^3} - \frac{k^2}{a^2H} - \frac{2 \rho_r}{H} -\frac{3 Q y^2}{2H},
\nonumber \\
&& a_{41}= 3 Q y + \frac{2 y\rho_r}{3H^2},
\nonumber\\
&& a_{42}=0,\; a_{43} = \frac{1}{3H}, \; a_{44} = 3 - \frac{2 \rho_r}{3H^2}.
\label{MatrixA}
\end{eqnarray}
The column matrix ${\bf B}$ has elements\footnote{In order to write the element $b_2$,
we make use of the result valid for Langevin equations with multiple noise terms~\cite{2noise1,2noise2}:
$\dot v = - \sum_i \gamma_i v + \sum_i n_i^{1/2} \xi \equiv
- \sum_i \gamma_i v + \sqrt{\sum_i n_i} \xi$, where $n_i$ are the amplitude of the noises. 
We have explicitly verified numerically, 
using an auxiliary Langevin code, that this identity holds.}:
\begin{eqnarray}
&&b_1 = 0,\; b_2 = \left[ \frac{6 Q T}{a^3 H^2} + \frac{(9+12\pi Q)^{1/2}(1+2n_*)}{\pi a^3 H}\right]^{1/2},
\nonumber \\
&& b_3 = - \frac{\left(6 Q T\right)^{1/2} y }{a^{3/2}},\; b_4=0.
\label{MatrixB}
\end{eqnarray}
{}Finally, the column matrix ${\bf C}$ appearing in the definition of the power spectrum
eq.~(\ref{PRPhi}) has the elements:
\begin{eqnarray}
&& c_1 = \frac{y H}{y^2 + 4 \rho_r/3}, \; c_2=0, \; c_3=0,\; c_4= -\frac{H}{y^2 + 4 \rho_r/3}.
\label{MatrixC}
\end{eqnarray}

\acknowledgments

R.O.R. acknowledges financial support by research grants from Conselho
Nacional de Desenvolvimento Cient\'{\i}fico e Tecnol\'ogico (CNPq),
Grant No. 307286/2021-5, and from Funda\c{c}\~ao Carlos Chagas Filho
de Amparo \`a Pesquisa do Estado do Rio de Janeiro (FAPERJ), Grant
No. E-26/201.150/2021. 
The work of G.S.R. is supported by a PhD scholarship from FAPERJ.
The authors acknowledge \texttt{Wolfram} for providing
a \texttt{Mathematica} for sites software license. 


\end{document}